\documentclass[journal, final]{IEEEtran}
\usepackage{amsmath}
\usepackage{amssymb}
\usepackage{amsthm}
\usepackage{graphicx}
\usepackage{graphics}
\usepackage{epstopdf}
\usepackage{algorithm}
\usepackage{algpseudocode}
\usepackage{multirow}
\usepackage{algorithmicx}
\usepackage{enumerate}
\usepackage{epstopdf}
\usepackage{cite}
\usepackage{amscd}
\usepackage{dsfont}
\usepackage{latexsym}
\usepackage{array}

\usepackage{tikz}\usetikzlibrary{calc}
\usepackage{tabularx}
\usepackage{epsfig}
\usepackage{graphicx,subfigure}
\usepackage{epstopdf}
\usepackage{graphicx}
\usepackage{amsmath,mathrsfs}
\usepackage{amsthm}
\usepackage{amssymb}  
\usepackage{amsbsy}
\usepackage{color}
\usepackage{stfloats}
\usepackage{algpseudocode}
\usepackage{bm}
\usepackage{pifont}

\newcommand\Algphase[1]{%

\Statex\hspace*{-\algorithmicindent}\textbf{#1}%

\algdef{SE}[SUBALG]{Indent}{EndIndent}{}{\algorithmicend\ }%
\algtext*{Indent}
\algtext*{EndIndent}
}

\ifCLASSINFOpdf

\else

\fi

\begin{document}

\title{Efficient Integer Coefficient Search for Compute-and-Forward}
	\author{William~Liu, 
	        Cong~Ling,~\IEEEmembership{Member,~IEEE}
	\thanks{William Liu and Cong Ling are with the Department of Electrical and Electronic Engineering,
	 Imperial College London, London, UK (e-mails: wl1007@imperial.ac.uk, cling@ieee.org).}
	\thanks{The work of William Liu is supported by the Engineering and Physical Sciences Research Council (EPSRC).}
}


\markboth{IEEE TRANSACTIONS on WIRELESS COMMUNICATIONS, TO APPEAR}%
{Shell \MakeLowercase{\textit{et al.}}: Bare Demo of IEEEtran.cls for Journals}

%




\maketitle

\begin{abstract}

Integer coefficient selection is an important decoding step in the implementation of compute-and-forward (C-F) relaying scheme. Choosing the optimal integer coefficients in C-F has been shown to be a shortest vector problem (SVP) which is known to be NP hard in its general form. Exhaustive search of the integer coefficients is only feasible in complexity for small number of users while approximation algorithms such as Lenstra-Lenstra-Lovasz (LLL) lattice reduction algorithm only find a vector within an exponential factor of the shortest vector. An optimal deterministic algorithm was proposed for C-F by Sahraei and Gastpar specifically for the real valued channel case. In this paper, we adapt their idea to the complex valued channel and propose an efficient search algorithm to find the optimal integer coefficient vectors over the ring of Gaussian integers and the ring of Eisenstein integers.  A second algorithm is then proposed that generalises our search algorithm to the Integer-Forcing MIMO C-F receiver. Performance and efficiency of the proposed algorithms are evaluated through simulations and theoretical analysis. 
\end{abstract}


\IEEEpeerreviewmaketitle

\section{Introduction}

\IEEEPARstart{I}{n} the past few years research activities in physical layer network coding have generated a huge amount of interest, in particular the compute-and-forward scheme proposed in\cite{nazer2011compute} which is shown to achieve vastly higher rates than conventional forms of network coding \cite{nazer2011compute}, \cite{wilson2010joint}. The underlying principle of C-F is that each relay should exploit the natural linear computations of the channel to decode an integer linear combination of transmitted messages from the source nodes, which are subsequently forwarded to the end nodes where the original transmitted messages can be solved in a straightforward manner provided the matrix of integer coefficients is of full rank. Finding the optimal integer coefficients at each relay to maximise the overall computation rate of C-F is therefore a key decoding step of the C-F scheme.

The C-F decoding at relay nodes in \cite{nazer2011compute} is facilitated by the use of nested lattice codes built from Construction A over the ring of integers, which ensures the integer linear combination of codewords also yields a codeword. C-F using lattices created using Construction A over the ring of Gaussian integers and the ring of Eisenstein integers are covered in \cite{feng2013algebraic} and \cite{tunali2012lattice}, \cite{sun2013lattice} respectively. It is shown in \cite{tunali2012lattice} that C-F with lattices over Eisenstein integers achieves superior performance in outage rate and error probability than C-F over Gaussian integers. An adaptive C-F scheme utilizing lattices  over rings of imaginary quadratic integers is given in \cite{huang2015adaptive}.

The most basic approach to finding the optimal integer coefficients in C-F is by peforming exhaustive search  \cite{nazer2011compute}, \cite{feng2013algebraic} over all integer vectors within a given range. Although exhaustive search guarantees the optimal integer coefficients everytime, this approach is practically non-viable due to a search complexity that grows exponentially with number of transmitting users. In \cite{feng2013algebraic}, the authors showed that finding the optimal integer coefficients in C-F is a Shortest Vector Problem (SVP), which is  known to be NP hard in its exact form. Therefore, lattice reduction based approximation algorithms have been proposed in \cite{feng2013algebraic} and \cite{sakzad2012ergodic} for the standard C-F scheme with single receive antenna, and \cite{sakzad2013integer} for the Integer-Forcing MIMO C-F receiver in \cite{zhan2014integer} and \cite{zhan2009mimo}. These algorithms use the basis of the reduced lattice to find a short vector within an exponential factor of the shortest length vector. Due to the suboptimality of the approach, using approximation algorithms such as the LLL \cite{lenstra1982factoring}  to find the optimal integer coefficients will result in significant computational rate loss in C-F that increases with the number of users.

A heuristic approach of finding integer coefficients called quantized exhaustive search(QES) was proposed in \cite{sakzad2014phase}  for the phase-precoded C-F introduced in the same work. With this method the relay iteratively adjusts the phase and magnitude of the C-F scaling factor to make the scaled channel coefficients as close as possible to the optimal integer coefficients.  However the authors do not explicitly give methods on how to select the size of the search step for phase and magnitude to guarantee the optimal solution, hence this method is considered suboptimal. The complexity of this algorithm depends on the search step size for both phase and magnitude of the the scaling factor, as well the absolute value of the maximum range for the scaling factor.

In a more recent work, the authors in \cite{sahraei2014compute} proposed an optimal integer coefficients search method for the case of real valued channels which shows that the SVP problem resulting from C-F can be solved exactly through an algorithm of low polynomial complexity in the number of transmitting users. An improvement was made on this work in \cite{sahraei2015polynomially} and \cite{wen2016linearithmic} to reduce the complexity of the search algorithm, although real valued channels were assumed to derive the reduced complexity in each case. Since we have to deal with complex valued gains for wireless channels in most real application scenarios, optimal coefficient search algorithms must be developed for complex valued channel cases.

The main contributions of this work is the following: building on the work of \cite{sahraei2014compute}, we extend their integer coefficient search algorithm for real valued channels to C-F over complex valued channels.  We propose two algorithms with low polynomial complexity in relation to the number of transmitting users, that find the exact solution for the complex valued version of the SVP in C-F.   Algorithm~\ref{alg:alg_1} finds the optimal equation coefficients for C-F with single receive antenna over the ring of Gaussian integers and the ring of Eisenstein integers.  Algorithm~\ref{alg:alg_2} finds the optimal equation coefficients for MIMO C-F over the ring of Gaussian integers and the ring of Eisenstein integers. The crux of both algorithms is Theorem 1 which specifies an exact relationship for the optimal solution, relating the optimal coefficients to the underlying channel coefficients. In comparison to the original algorithm given for C-F over real valued channel, we make necessary modifications to streamline our proposed algorithms in order to make it efficient for computations over the complex valued channels. The complexity of our proposed algorithms is examined in detail and simulation analysis is provided to compare the performance and efficiency of Algorithm 1 with other efficient integer coefficient search algorithms.

\subsection{Notation}

In this paper we use the following notations. $\mathbb{Z}$, $\mathbb{Z}\left [ j \right ]$ and $\mathbb{Z}\left [ \omega \right ]$ denotes the set of integers, Gaussian integers and Eisenstein integers respectively, where $\omega =-\frac{1}{2}+j\frac{\sqrt{3}}{2}$. $ \mathbb{R}$ and $\mathbb{C}$ represent the set of real and complex numbers respectively. For $x\in \mathbb{R}$, we denote $\left[x\right]$ as rounding to the nearest integer with half integers rounded upwards.  For $x\in \mathbb{C}$ we denote $\left [ x \right ]_{\mathbb{Z}\left [ j  \right ]}$ as quantizing to the nearest Gaussian integer, whereby the real and imaginary components of $x$ are individually rounded to the nearest integer, with half integers rounded upwards. We denote  $\left [ x \right ]_{\mathbb{Z}\left [ \omega  \right ]}$ as quantizing to the nearest Eisenstein Integer, an operation which can be done using an $A_{2}$ lattice decoder \cite{conway1982fast}, \cite{mow1992fast}. We denote $\left |x  \right |$ as the absolute value  of a real number $x\in \mathbb{R}$, and as the modulus of a complex number $x\in \mathbb{C}$. We denote $\left \|\cdot   \right \|$ as the Euclidean norm of a real or complex vector. Scalars are denoted by plain font and vectors by bold lower case font. Bold capital letters represent matrices. For a matrix $\mathbf{H}$, the element $H_{mn}$ refers to the element in row $m$ and column $n$. For a vector $\mathbf{h}$ we denote the $n$-th element by $h_{n}$. When denoting indexed vectors, $\mathbf{h}_{m}$ indicates the $m$-th row vector and $\mathbf{h}_{n}^{T}$ the $n$-th column vector of $\mathbf{H}$. For an $k \times L$ matrix $\mathbf{H}$, notation $\mathbf{H}_{\tau }$ denotes a submatrix of $\mathbf{H}$ consisting of columns indexed in the set $\tau\subseteq \left \{1,...,L \right \}$. 

\section{Problem Formulation and Existing Search Methods for Compute-and-Forward}

For a general relay network of multiple source nodes transmitting to multiple relay nodes, we consider the operations from the perspective of a single relay node with a single receive antenna. The messages $\mathbf{w}_{1},\mathbf{w}_{2},\cdots ,\mathbf{w}_{L}\in \mathbb{F}_{q}^{k}$ from the $L$ source nodes are first encoded to $L$ number of $n$-dimensional complex valued codewords  $\mathbf{x}_{1}, \mathbf{x}_{2},\cdots ,\mathbf{x}_{L}\in \mathbb{C}^{n}$ which are then transmitted over channel with complex-valued coefficients $h_{l}$ for $l=0,..,L$. Denote $\Lambda_{1}$ as a fine lattice and $\Lambda$ as a coarse lattice, the codewords to be sent are drawn from a nested lattice code defined by the quotient $\Lambda_{1} /\Lambda$ where $\Lambda \subseteq \Lambda_{1}$. The received signal at each relay node is a linear combination of the transmitted signals from all the source nodes plus additive noise which can be expressed as, 

\begin{equation}\label{Eq1}
\mathbf{y}=\sum_{l=1}^{L}h_{l}\mathbf{x}_{l}+\mathbf{z}.
\end{equation}

\noindent The fading channel coefficients between the source nodes and the relay nodes are assumed perfectly known at the relay nodes. The $\mathit{n}$-dimensional codewords are subjected to average power constraint of $\frac{1}{\mathit{n}}\mathit{E}\left \| \mathbf{x}_{l} \right \|^{2}\leq \mathit{P}$ and $\mathbf{z}\sim \mathbb{C}\textup{N}\left ( 0,\sigma ^{2}\mathbf{I} \right )$ is a circularly-symmetric jointly-Gaussian complex random vector where $\mathbf{I}$ is the $n\times n$ identity matrix. 

The target of the C-F decoder at each relay is to  decode from the noisy linear signal in (\ref{Eq1}), an estimate $\hat{\mathbf{u}}$ of an integer linear combination of the transmitted codewords given by,

\begin{equation}\label{Eq2}
\mathbf{u}=\left [ \sum_{l=1}^{L}a_{l}\mathbf{x}_{l} \right ]\textup{mod }\Lambda, 
\end{equation}

\noindent where $a_{l}$ is the  optimal  integer coefficient to be determined as part of the decoding process, and $\Lambda$ is the C-F shaping lattice. The nature of the integer coefficients depends on the underlying channel and takes on the form of rational integers for real valued channels and Gaussian integers for complex valued channels. A C-F scheme over Eisenstein integers is also proposed in [4].

The first step of C-F decoding is to scale the received signal by the minimum mean square error (MMSE) coefficient. The scaled signal,
\begin{equation}\label{Eq3}
\tilde{\mathbf{y}}=\alpha \mathbf{y}=\sum_{l=1}^{L}a_{l}\mathbf{x}_{l}+\mathbf{z}_\textup{eff},
\end{equation}
\noindent can be expressed in terms of the desired integer linear combination of transmitted codewords and the effective noise $\mathbf{z}_\textup{eff}$, which is a result of channel mismatch and scaling of original noise,

\begin{equation}\label{Eq4}
\mathbf{z}_\textup{eff}=\sum_{l=1}^{L}\left ( \alpha h_{l}-a_{l} \right )\mathbf{x}_{l}+\alpha \mathbf{z}.
\end{equation}

\noindent Standard lattice decoder denoted by $Q_{\Lambda_{1} }(.)$ is then used to quantize the scaled signal to the nearest fine lattice point in $\Lambda_{1}$. The quantized signal is then modulo operated over the shaping lattice $\Lambda$ to decode an estimate $\hat{\mathbf{u}}$ of the integer linear combination. 

\begin{equation}\label{Eq5}
\hat{\mathbf{u}}=Q_{\Lambda_{1} }\left ( \alpha \mathbf{y} \right )\textup{mod }\Lambda. 
\end{equation}

\noindent The optimal coefficient vector $\mathbf{a}=\left [ a_{1}, a_{2},\cdots ,a_{L} \right ]\in  \mathbb{Z}[j]^{L}$ is selected to maximise the achievable computational rate of C-F at the relay, which is proven in \cite{nazer2011compute}  to be,

\begin{equation} \label{Eq6}
R(\mathbf{h},\mathbf{a})=\mathop{\max }\limits_{\alpha \in \mathbb{C}} \log^{+}  \left(\frac{P}{\left|\alpha \right|^{2} +P\left\| \alpha \mathbf{h}-\mathbf{a}\right\| ^{2} }\right).
\end{equation}

\noindent It is further shown in \cite{nazer2011compute} that this computation rate can be uniquely maximized by further choosing $\alpha$ to be the MMSE coefficient which is given by, 

\begin{equation} \label{Eq7}
\alpha _{\textup{MMSE}} =\frac{\mathbf{a}\mathbf{h}^{H} P}{1+P\left\| \mathbf{h}\right\| ^{2} }.
\end{equation}

\noindent By substituting MMSE scaling factor $\alpha _{\textup{MMSE}}$ into computational rate in (\ref{Eq6}) yields,
\begin{equation}\label{Eq8}
R\left ( \mathbf{h},\mathbf{a} \right )=\textup{log }^{+}\left (\left (\left \|\mathbf{a}  \right \|^{2}-
\frac{P\left | \mathbf{a}\mathbf{h}^{H}\right |^{2}}{1+P\left \| \mathbf{h} \right \|^{2}}  \right )^{-1}  \right ).
\end{equation}

\noindent In \cite{nazer2011compute} it is also shown that the computational rate is greater than zero for the range,

\begin{equation}\label{Eq9}
\left \| \mathbf{a} \right \|\leq \Phi =\sqrt{1+P\left \| \mathbf{h} \right \|^{2}}.
\end{equation}

\noindent Rate optimization of C-F can be equivalently expressed as a minimization problem on the effective noise of C-F in (\ref{Eq4}). Substituting the MMSE coefficient into the effective noise term, we obtain the cost function $f\left ( \mathbf{a}\right )$ to be minimised in (\ref{Eq10}). Therefore for C-F over complex valued channels, finding the optimal equation coefficients can be re-expressed as finding the Gaussian integer vector $\mathbf{a}\in  \mathbb{Z}[j]^{L}$ that minimizes the target cost function $f\left ( \mathbf{a}\right )$,
\begin{equation} \label{Eq10}
\begin{split}
\mathbf{a}^{\textup{opt}}=\underset{\mathbf{a}\in  \mathbb{Z}[j]^{L}\setminus \mathbf{a}= \mathbf{\mathbf{0}}}{\operatorname{argmin}}f\left ( \mathbf{a} \right )
&=\underset{\mathbf{a}\in  \mathbb{Z}[j]^{L}\setminus \mathbf{a}= \mathbf{0}}{\operatorname{argmin}}\left(\textbf{a}\mathbf{M}\mathbf{a}^{H}  \right )\\
&=\underset{\mathbf{a}\in  \mathbb{Z}[j]^{L}\setminus \mathbf{a}= \mathbf{0}}{\operatorname{argmin}}\left \| \mathbf{a}\mathbf{L}  \right \|^{2},
\end{split}
\end{equation}

\noindent where,

\begin{equation}\label{Eq11}
\mathbf{M}=\left (1+P\left \| \mathbf{h} \right \|^{2}  \right )\cdot \mathbf{I}-P\cdot \mathbf{h}^{H}\mathbf{h}.
\end{equation}

\noindent $\mathbf{M}$ is a Hermitian positive definite matrix and has a Cholesky decomposition $\mathbf{M}=\mathbf{L}\mathbf{L}^{H}$, where $\mathbf{L}$ is a lower triangular matrix. 
Therefore finding the optimal integer coefficients in C-F can be viewed as the SVP of finding the non-zero minimal Euclidean norm point in a lattice with Gram matrix $\mathbf{M}$. 

In \cite{zhan2014integer} and \cite{zhan2009mimo}, MIMO C-F was introduced where the relay aims to decode the best integer linear combination of codewords using multiple antennas. The MIMO C-F system model consists of $L$ transmitting users sending to relays with $k$ receive antennas. Take $\mathbf{H}$ to be a $k \times L$ channel matrix and let $\mathbf{b}=\left [ b_{1}, b_{2},\cdots ,b_{k} \right ] \in \mathbb{C}^{k}$ be a judiciously chosen preprocessing vector which is applied to the channel output vector $\mathbf{y}$  at each relay. The relay aims to select the optimal coefficient vector $\mathbf{a}=\left [ a_{1}, a_{2},\cdots ,a_{L} \right ]\in \mathbb{Z}\left [ j \right ]^{L}$ to maximise the achievable computational rate of MIMO C-F, which is proven in \cite{zhan2014integer}  to be,

\begin{equation}\label{Eq12}
R\left ( \mathbf{H},\mathbf{a},\mathbf{b} \right )=\frac{1}{2}\textup{log}^{+}\left ( \frac{P}{\left \| \mathbf{b} \right \|^{2}+P\left \| \mathbf{b}\mathbf{H}-\mathbf{a} \right \|^{2}  } \right ).
\end{equation}

\noindent From the perspective of a single relay with multiple receive antennas, it is also shown in \cite{zhan2014integer} that the optimal integer forcing vector $\mathbf{b}$ that maximises the achievable rate in (\ref{Eq12}) is given by,

\begin{equation}\label{Eq13}
\mathbf{b}_{\textup{opt}}=\mathbf{a}\mathbf{H}^{H}\left ( P^{-1}\mathbf{I}+\mathbf{H}\mathbf{H}^{H}\right )^{-1}
\end{equation}

\noindent To maximize the computational rate (\ref{Eq12}), it is shown in \cite{zhan2014integer} that it is sufficient to check the space of all Gaussian integer vectors $\mathbf{a}$ satisfying,

\begin{equation}\label{Eq14}
\left \| \mathbf{a} \right \|\leq \Phi =\sqrt{1+P\lambda_{max}^{2}},
\end{equation}

\noindent where $\lambda_{max}$  is the maximum singular value of $\mathbf{H}$. Substituting optimal integer forcing vector $\mathbf{b}_{\textup{opt}}$ into (\ref{Eq12}),  the rate optimization problem can be expressed as,

\begin{equation} \label{Eq15}
\begin{split}
\mathbf{a}^{\textup{opt}}=\underset{\mathbf{a}\in  \mathbb{Z}[j]^{L}\setminus \mathbf{a}= \mathbf{0}}{\operatorname{argmin}}f\left ( \mathbf{a} \right )
&=\underset{\mathbf{a}\in  \mathbb{Z}[j]^{L}\setminus \mathbf{a}= \mathbf{0}}{\operatorname{argmin}}\left(\textbf{a}\mathbf{M}\mathbf{a}^{H}  \right )\\
&=\underset{\mathbf{a}\in  \mathbb{Z}[j]^{L}\setminus \mathbf{a}= \mathbf{0}}{\operatorname{argmin}}\left(\mathbf{a}\mathbf{V}\mathbf{D}\mathbf{V}^{H}\mathbf{a}^{H}  \right ),
\end{split}
\end{equation}

\noindent where $\mathbf{V}\in \mathbb{C}^{L\times L}$ is composed of the right singular vectors of $\mathbf{H}$ and $\mathbf{D}\in \mathbb{C}^{L\times L}$ is a diagonal matrix whose first $k$ diagonal elements satisfy $r_{i}=\left (1+P\lambda_{i}^{2}  \right )^{-1},  i=1,...,k$ and the remaining $L-k$ elements equal to 1. $\lambda_{i}$ is the $i$-th singular value of $\mathbf{H}$. Therefore the search for the optimal integer coefficients vector for MIMO C-F is equivalent to finding the shortest vector in a lattice with gram matrix $\mathbf{V}\mathbf{D}\mathbf{V}^{H}$.

\subsection{Overview of Exisiting Coefficient Search Methods}

In this section, we review in detail the main integer coefficient search methods for C-F proposed in literature and discuss their performance and complexity.

1). Exhaustive search of integer coefficients \cite{nazer2011compute}: The optimal integer coefficients in C-F can be found through exhaustive search by checking all the non-zero integer vectors of length $L$ in a sphere of radius given by the bound on the norm of integer coefficients in (\ref{Eq9}) for standard C-F and (\ref{Eq14}) for MIMO C-F. The complexity of exhaustive search is given by approximately $O(\Phi^{2L} )$ where $\Phi=\sqrt{1+P\left \| \mathbf{h}^{2} \right \|}$. Since the complexity increases exponentially with the number of transmitting users, this method has the highest computational complexity, although it guarantees the optimal solution. 

2). Lattice reduction based algorithms focused around the LLL algorithm \cite{feng2013algebraic}, \cite{sakzad2012ergodic}, \cite{lenstra1982factoring} for standard C-F. For the SVP in Integer-Forcing MIMO C-F, proposed lattice reduction based algorithms include complex-LLL(CLLL) \cite{sakzad2013integer}, \cite{gan2009complex}, Hermite-Korkine-Zolotareff(HKZ)\cite{sakzad2013integer} and Minkowski reduction algorithm\cite{sakzad2013integer}. The underlying principle of the LLL algorithm, which has a complexity of $O\left (  L^{4}\textup{log}L\right )$, is to compute a reduced basis from the original basis of the lattice, where the first vector of the LLL-reduced basis approximates the shortest vector of the lattice. For $L$ users the length of the LLL-approximated shortest vector is at most $2^{\frac{L-1}{2}}$ times the length of the shortest vector in the lattice. Therefore as $L$ increases, the rate loss that occurs from using LLL for C-F becomes significant enough such that an optimal yet computationally efficient search method is required. For SVP over MIMO C-F, the proposed HKZ and Minkowski reduction algorithms in \cite{sakzad2013integer} with complexity of $\left ( 2\pi e \right )^{L+O\left ( \textup{log}2L \right )}$ and $\left ( 5/4 \right )^{2L^{2}}$ respectively, offer improved performances over CLLL at the cost of higher complexity for higher order MIMO channels. 
 
3). Quantized exhaustive search [8] is a streamlined version of the standard exhaustive search method which is originally proposed for the phase precoded C-F, but is also relevantly applicable to standard C-F scheme. The basis of the method uses the inferred relationship between optimal integer coefficients vectors and channel coefficients,
\begin{equation}\label{Eq16}
\mathbf{a} =\left [ \alpha \mathbf{h} \right ]_{\mathbb{Z}\left [ j \right ]},
\end{equation}
\noindent whereby $\alpha \in \mathbb{C}$. This relationship was not proven in \cite{sakzad2014phase} but made based on heuristic analysis of the C-F model. In quantized exhaustive search, the complex scaling factor $\alpha$ is broken down into its phase and magnitude components. The algorithm changes the magnitude component from 0 to the maximum magnitude $\left |\alpha   \right |_{\textup{max}}$ in steps of $\left |\alpha   \right |_{\textup{step}}$, while simultaneously changing the the phase component of $\alpha$ from 0 to 90 in steps of $d$. The $\alpha$ value taken at  each iteration is used to make an initial estimate of integer coefficient vector $\mathbf{a}$ using (\ref{Eq16}), which is then used to calculate the corresponding MMSE coefficient (\ref{Eq7}). The estimated $\mathbf{a}$ along with its corresponding MMSE coefficient is then used to evaluate the effective noise term of C-F in (\ref{Eq4}), and the $\mathbf{a}$ that minimises the effective noise is taken as the optimal integer coefficient vector. The complexity of this algorithm is shown to be $O(90/d\times \left | \alpha \right |_{\textup{max}}/\left |\alpha   \right |_{\textup{step}})$. Whether this algorithm can yield the optimal integer coefficients depends entirely on selecting $d$ and $\left |\alpha   \right |_{\textup{step}}$ to be sufficiently small for a specific set of channel instances. Since the authors do not outline explicitly how to select the exact search parameters to guarantee the optimal coefficients, this search method can be considered suboptimal. 

4). An optimal integer coefficient search algorithm of polynomial complexity was proposed in \cite{sahraei2014compute} for C-F over real valued channels. The authors proved that the solution to minimization problem of (\ref{Eq10}) for integer valued coefficients,  satisifies,
\begin{equation}\label{Eq17}
\mathbf{a}^{\textup{opt}}=\left [ \alpha \mathbf{h} \right ],
\end{equation}
\noindent for $\alpha \in \mathbb{R}$, otherwise  $\mathbf{a}^{\textup{opt}}$ must be a standard unit norm vector. This is the fundamental theorem on which the optimal polynomial complexity algorithm in \cite{sahraei2014compute} is based around. Using (\ref{Eq17}), the authors showed that finding the optimal integer coefficients becomes a one-dimensional optimization problem where the cost function $f\left ( \mathbf{a}\left ( \alpha  \right ) \right )$ to be minimised in (\ref{Eq9}) is a piece-wise constant function in $\mathbf{a}\left ( \alpha  \right )$ and therefore a piecewise constant function in $\alpha $. The authors find a set of points of $\alpha$ that represent all intervals in which $f\left ( \mathbf{a}\left ( \alpha  \right ) \right )$ is constant and checks the value of $f\left ( \mathbf{a}\left ( \alpha  \right ) \right )$ over all points. This is done by first finding all the discontinuity points of $a_{l}\left ( \alpha  \right )=\left [ \alpha h_{l} \right ]$, for $l=1,...,L$, which occur at points where $\alpha h_{l} $ is a half integer. Then, using the discontinuity points of $a_{l}\left ( \alpha  \right )$ the authors proceed to find all the $\alpha$ values for  which  $a_{l}\left ( \alpha  \right )$ is discontinuous, which are also the $\alpha$ values for which the function $f\left (\mathbf{a}\left ( \alpha  \right ) \right )$ is discontinuous. The $\alpha$ discontinuities points are then sorted in ascending order and the middle value between consecutive discontinuities are used to evaluate $ f\left ( \mathbf{a}\left ( \alpha  \right ) \right )$. The $\alpha$ value that minimises the target function therefore generates the optimal integer coefficients based on relationship (\ref{Eq17}). The complexity of this algorithm is shown to be $O\left (L^{2}\Phi  \right )$. An improvement is made on this algorithm by the original authors in \cite{sahraei2015polynomially}  to reduce the complexity to $O\left ( L\Phi \log \left ( L\Phi \right ) \right )$. In \cite{wen2016linearithmic}, a further improvement is proposed to reduce the expected complexity to $O\left ( L \log L \right )$. However, as with \cite{sahraei2014compute} and \cite{sahraei2015polynomially} , the work of \cite{wen2016linearithmic} assumes real valued channels to arrive at this result.

\section{Proposed Optimal algorithm for Complex valued channel}

Since the channel coefficients to be quantised as part of C-F decoding lies in $\mathbb{C}$, it is natural that we should first consider the case where the integer coefficients decoded at the relay belong to the ring of Gaussian integers. For a C-F system of $L$ transmitting users sending to relays with a single receive antenna over complex channels $\mathbf{h}\in\mathbb{C}^{L}$, we first propose an algorithm that finds the optimal coefficient vector $\mathbf{a}\in \mathbb{Z}\left [ j \right ]^{L}$ for C-F over Gaussian integers. Additionally, since it is known that $\mathbb{Z}[\omega]$ quantizes $\mathbb{C}$ best among all imaginary quadratic integers, we also adapt our algorithm to find the optimal coefficients $\mathbf{a}\in \mathbb{Z}\left [ \omega \right ]^{L}$ for C-F over Eisenstein integers. We then propose a search algorithm for MIMO C-F of $L$ transmitting users and $k$ receive antennas at the relay, that finds the optimal coefficient vector in both $\mathbb{Z}\left [ j \right ]$ and $\mathbb{Z}\left [ \omega \right ]$. The basis of our two algorithms is Theorem 1  which shows that the SVP problem in its complex form in C-F has an exact solution. 

\newtheorem{theorem}{Theorem}
\begin{theorem}
\label{theorem:1}
\noindent  For C-F over complex ring $\mathrm{R}$, where $\mathrm{R}$ denotes either the ring of Gaussian integers or the ring of Eisenstein integers, the solution to (\ref{Eq14}) satisfies,
\begin{equation}\label{Eq18}
\mathbf{a}^{\textup{opt}}=\left [  \mathbf{b}\mathbf{H} \right ]_{\textup{R}}=\begin{cases}
  \mathbf{a}^{\textup{opt}}=\left [    \mathbf{b}\mathbf{H}  \right ]_{\mathbb{Z}\left [ j \right ]},  \text{ for } \mathbf{a}\in  \mathbb{Z}[j]^{L},  \\ 
  \mathbf{a}^{\textup{opt}}=\left [   \mathbf{b}\mathbf{H}   \right ]_{\mathbb{Z}\left [ \omega \right ]},  \text{ for } \mathbf{a}\in  \mathbb{Z}[\omega]^{L}.
\end{cases}
\end{equation}
\noindent for some $\mathbf{b}\in \mathbb{C}^{k}$, or $\mathbf{a}^{\textup{opt}}$ must be a unit norm vector in the ring $\mathrm{R}$. 
\end{theorem}
The unit norm vector refers to any vector where every element of $\mathbf{a}^{\textup{opt}}$ is zero with the exception of one element, with the nonzero element taking the value of $1$, $-1$, $j$ or $-j$ for the ring of Gaussian integers, or $1$, $-1$, $\pm \omega$, $\pm \omega^{2}$ for the ring of Eisenstein integers.

\begin{IEEEproof}
We first give proof for the case $\mathbf{a}\in \mathbb{Z}\left [ j \right ]^{L}$. Consider the optimization problem of finding $\mathbf{a}\in \mathbb{Z}\left [ j \right ]^{L} \setminus \mathbf{0}$ that maximizes the computational rate of MIMO C-F in (\ref{Eq12}). This is clearly equivalent to the minimization problem on the target cost function $f(\mathbf{a})$ in (\ref{Eq15}). Let $\left ( \mathbf{b}^{\textup{opt}}, \mathbf{a}^{\textup{opt}} \right )$ be the optimal solution and let $\mathbf{h}_{n}^{T}$ be the $n$-th column vector of $\mathbf{H}\in \mathbb{C}^{k\times L}$. The two possible cases are, for $n=0,1,.....,L$,

\begin{enumerate}
\item $\forall n:\left [ \mathbf{b}^{\textup{opt}}\mathbf{h}_{n}^{T} \right ]_{\mathbb{Z}\left [ j \right ]}=0$, in this case $\mathbf{a}^{\textup{opt}}$ is a Gaussian integer vector of unit norm,
\item $\exists n:  \left [ \mathbf{b}^{\textup{opt}}\mathbf{h}_{n}^{T} \right ]_{\mathbb{Z}\left [ j \right ]}\neq 0$, in which case $\mathbf{a}^{\textup{opt}}=\left [\mathbf{b}^{\textup{opt}}\mathbf{H} \right ]_{\mathbb{Z}\left [ j \right ]}$.
\end{enumerate}

\noindent Therefore the optimal solution takes on the form of $\mathbf{a}^{\textup{opt}}=\left [\mathbf{b}^{\textup{opt}}\mathbf{H} \right ]_{\mathbb{Z}\left [ j \right ]}$ or it will be a Gaussian integer vector of unit norm. The proof for Theorem 1 over the ring of  Eisenstein integers can be deduced in a similar manner. 
\end{IEEEproof}

\subsection{Proposed Algorithm: C-F over Gaussian Integers}

We first present our algorithm for C-F over Gaussian integers for the complex channel, when the relay has one receive antenna. Note that for single receive antenna case of $k=1$, it can be deduced from Theorem 1 that the optimal coefficient vector will either take on the form of $\mathbf{a}^{\textup{opt}}=\left [   \alpha \mathbf{h}   \right ]_{\mathbb{Z}\left [ j \right ]}$ for some $\alpha \in \mathbb{C}$, or it will be a Gaussian integer vector of unit norm. Based on this relationship between optimal coefficient vector and $\alpha$, we can likewise express the target minimization function in (\ref{Eq10}) as a function of $\alpha$.

The first step in our search method is finding the set of values in $\alpha$ representing the discontinuity points of the continuous function $f\left (\mathbf{a}\left ( \alpha  \right ) \right )$, based on the discontinuity points of neighbouring Gaussian integer coefficients $\mathbf{a}\left ( \alpha  \right )$. This is possible since the values of $\alpha $ for which $\mathbf{a}\left ( \alpha  \right )\in \mathbb{Z}\left [ j \right ]^{L}$  is discontinuous are also values for which $f\left ( \mathbf{a}\left ( \alpha  \right ) \right )$ is discontinuous. Using the points of discontinuity for $f\left (\mathbf{a}\left ( \alpha  \right ) \right )$, we can then find a point to represent every constant interval of $f\left (\mathbf{a}\left ( \alpha  \right ) \right )$, therefore allowing us to evaluate the specific value of $\alpha$ and the corresponding $\mathbf{a}\left ( \alpha  \right )$ that minimises $f\left (\mathbf{a}\left ( \alpha  \right ) \right )$. Note that $f\left (\mathbf{a}\left ( \alpha  \right ) \right )$ is a two dimensional piecewise linear function due to the real and complex components of Gaussian integer variable.

From the minimization cost function in (\ref{Eq10}), it can be observed that the discontinuities in $f\left (\mathbf{a}\left ( \alpha  \right ) \right )$ occur when the Gaussian integer coefficients $\mathbf{a}\left ( \alpha  \right )\in \mathbb{Z}\left [ j \right ]^{L}$ change in value, or equivalently when its individual elements $a_{l}$ change in value, for $l=1,...,L$.  Since we know from Theorem~1 that $a_{l}\left ( \alpha  \right )=\left [ \alpha h_{l} \right ]$, the discontinuities of $a_{l}\left ( \alpha  \right )$ occur when the real part or the imaginary part, or both components of $\alpha h_{l}$ are half integers. From the rounding relationship denoted in Theorem~1, we can see that the significance of the discontinuities between neighbouring Gaussian integers is that they represent the exact cut-off points between rounding to one Gaussian integer or to the other, and therefore they also represent the discontinuities of $f\left (\mathbf{a}\left ( \alpha  \right ) \right )$. The discontinuities of Gaussian integer vector $\mathbf{a}\left ( \alpha  \right )$ that we need to consider are therefore the intermediate points between neighbouring Gaussian integer points in the real and imaginary domain. Denoting discontinuities of $\mathbf{a}\left ( \alpha  \right )$ by $\varphi$, these are complex valued points of the form,

\begin{equation}\label{Eq19}
\varphi \in \Psi :\left\{\begin{matrix}c+dj \; \textup{where} \;c\in \mathbb{Z}, \:d-\frac{1}{2}\in \mathbb{Z},
\\ \:e+fj \; \textup{where} \; e-\frac{1}{2}\in \mathbb{Z}, \:f \in \mathbb{Z},
\\ \qquad o+pj \;\textup{where} \; o-\frac{1}{2}\in \mathbb{Z}, \:p-\frac{1}{2}\in \mathbb{Z}.
\end{matrix}\right.
\end{equation}

\noindent Given the constraint on the optimal coefficient vector in (\ref{Eq9}) we know that it is only necessary to check discontinuities of $\mathbf{a}\left ( \alpha  \right )$ within a sphere of radius slightly larger than the bound in (\ref{Eq9}), namely,

\begin{equation}\label{Eq20}
\left | \mathbf{\varphi } \right |\leq \left \lceil \Phi \right \rceil+\frac{1}{2}.
\end{equation}

Using the relationship between $\mathbf{a}\left ( \alpha  \right )$ and $\alpha$ inferred in Theorem 1, along with the discontinuities of $\mathbf{a}\left ( \alpha  \right )$  that satisfy the bound in (\ref{Eq20}), we can deduce that the values of  $\alpha$ for which $\mathbf{a}\left ( \alpha  \right )$ and $f\left ( \mathbf{a}\left ( \alpha  \right ) \right )$ is discontinuous, are given by,
\begin{equation}\label{Eq21}
\Omega = \left \{\frac{\varphi }{h_{l}}\mid l=1,...,L, \, h_{l}\neq 0, \,  \left | \mathbf{\varphi } \right |\leq \left \lceil \Phi \right \rceil+\frac{1}{2}  \right \}.
\end{equation}

\begin{figure}[t] 
\centering
\includegraphics[width=8cm]{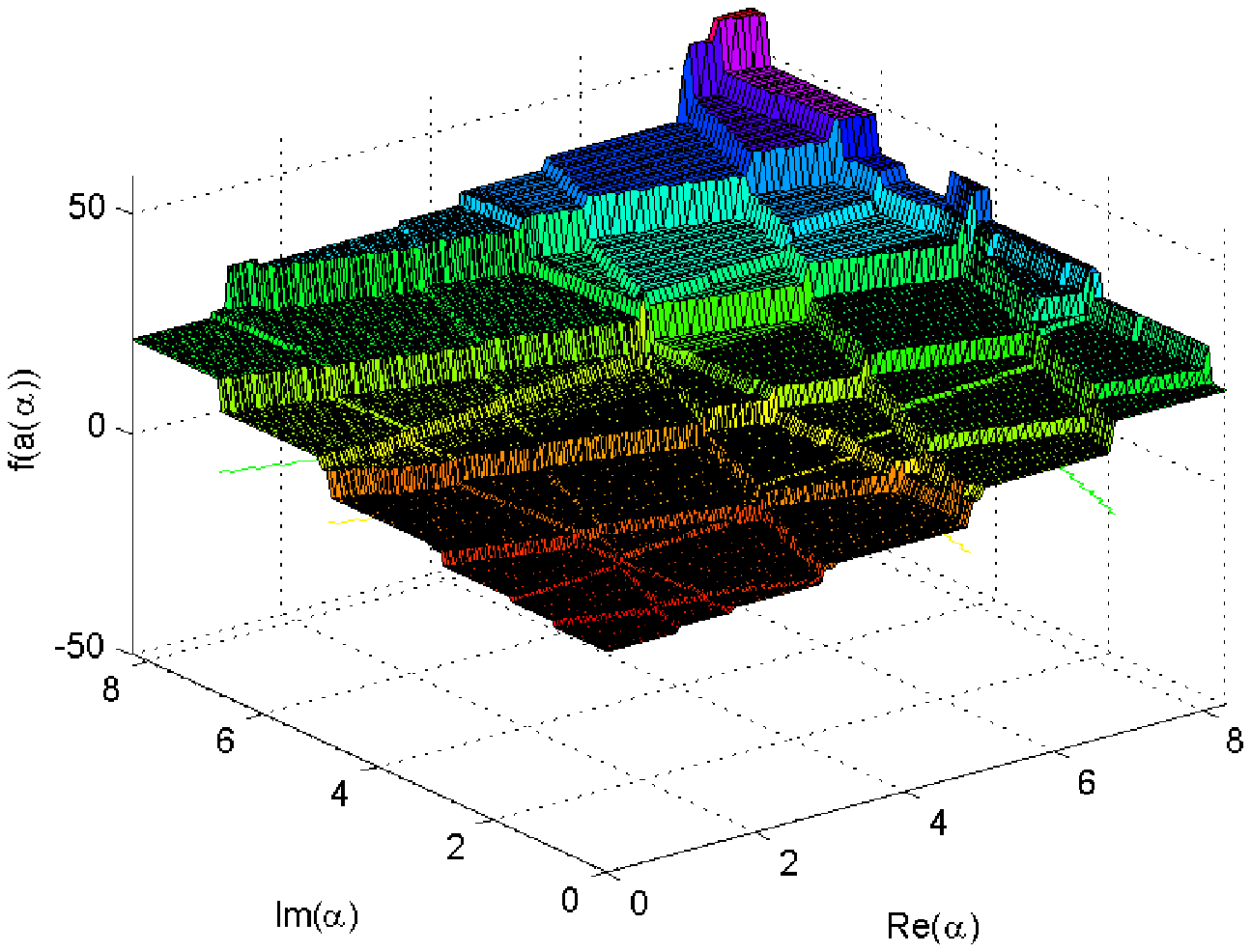}
\caption{The plot of target minization function $f\left ( \mathbf{a}\left ( \alpha  \right ) \right )$ against real and imaginary components of $\alpha$.}
\label{fig:1}
\end{figure}

\noindent  The discontinuities of $f\left ( \mathbf{a}\left ( \alpha  \right ) \right )$ given by the set $\Omega$ therefore form the vertices of planes where $f\left ( \mathbf{a}\left ( \alpha  \right ) \right )$ is locally constant. This can be illustrated by the three dimensional plot of the target function $f\left ( \mathbf{a}\left ( \alpha  \right ) \right )$ against the real and imaginary components of $\alpha$, shown in Figure~\ref{fig:1}, where each locally constant part of the function is composed of polytopes. To evaluate the minimization problem completely, we must find a value of $\alpha$ to represent every constant part of the function.  

Note that for the one dimensional piecewise constant function that occurs in the optimization problem for C-F over real valued channels, the authors in \cite{sahraei2014compute} proposed that the obtained discontinuities should be sorted and the average value of consecutive discontinuties used to evaluate $f\left ( \mathbf{a}\left ( \alpha  \right ) \right )$. If we were to translate this approach to the complex channel, then we need to find a point of $\alpha$ to represent each constant polytope plane of  $f\left ( \mathbf{a}\left ( \alpha  \right ) \right )$ and evaluate $f\left ( \mathbf{a}\left ( \alpha  \right ) \right )$ for that point. This optimization problem over multiple dimensions was tackled for the MIMO C-F model in \cite{sahraei2014compute} whereby when the number of receive antennas at the relay is more than one, the cost function to be minimised can also be viewed as a multi-dimensional piecewise linear function that can be solved by ensuring every constant polytope region of $f\left ( \mathbf{a}\left ( \alpha  \right ) \right )$ is checked. The authors in \cite{sahraei2014compute} proposed to find the average of every possible pairwise combinations of all the discontinuities in $\alpha$, which will guarantee that each constant region will have at least one representative. The problem with obtaining average values in this manner is that it makes this part of the algorithm inefficient, resulting in an excessively high complexity that depends on the total number discontinuities, which is determined by the bound in (\ref{Eq20}) and the number of transmitting users.

However, in our proposed algorithm we make an efficient simplification of the approach in \cite{sahraei2014compute} to significantly reduce computational complexity without sacrificing performance. In fact, it is sufficient to use the values of $\alpha$ for which $f\left ( \mathbf{a}\left ( \alpha  \right ) \right )$ is discontinuous, to directly evaluate the cost function in (\ref{Eq10}). For the real channel case, this is true because the piece-wise constant function will be open at one end and closed at the other, by the virtue of the rounding operation. When we evaluate $f\left ( \mathbf{a}\left ( \alpha  \right ) \right )$ directly using its discontinuities in $\alpha$, there will be a singular elements of  $\alpha \mathbf{h}$ that are exactly half integers. This means the value of $f\left ( \mathbf{a}\left ( \alpha  \right ) \right )$ for that $\alpha$ will be calculated using the $\mathbf{a}\left ( \alpha  \right )$ obtained by rounding up $\alpha \mathbf{h}$ at its half integer components. Therefore the discontinuity points can be used to obtain the value of $f\left ( \mathbf{a}\left ( \alpha  \right ) \right )$ for the particular interval which the discontinuity proceeds in the piecewise step function. Taking into account that $f\left ( \mathbf{a}\left ( \alpha  \right ) \right )$ is zero when $\alpha$ is zero, no interval is unaccounted for.

Therefore for the complex channel case, we can use the discontinuities of  $f\left ( \mathbf{a}\left ( \alpha  \right ) \right )$ denoted by set $\Omega$ to directly evaluate $f\left ( \mathbf{a}\left ( \alpha  \right ) \right )$. Note that the value of $\mathbf{a}(\alpha)$ calculated using Theorem 1  for $\alpha\in \Omega$ will assume the rounded up value of $\alpha\mathbf{h}$, where half integer values in real and imaginary planes are both rounded upwards. In this way, every discontinuity in $\alpha$ represents a plane where $f\left ( \mathbf{a}\left ( \alpha  \right ) \right )$ is constant. The minimization problem can be written as,
\begin{equation} \label{Eq22}
\mathbf{a}^{\textup{opt}}=\underset{\mathbf{a}\in  \mathbb{Z}[j]^{L}\setminus \mathbf{a}= \mathbf{0}}{\operatorname{argmin}}f\left ( \mathbf{a} \right )=\underset{\alpha\in \Omega}{\operatorname{argmin}} \, f\left ( \mathbf{a}\left ( \alpha\right )\right ).\\
\end{equation}

From an implementation perspective, there are a number of ways to optimise the search algorithm. One way to do this is to efficiently generate $\Psi$ in (\ref{Eq19}) and consequently $\Omega$ in (\ref{Eq21}). For small values of $L$, $\Omega$ can be easily generated by iterating over two loops, each loop running from   $-\left ( \left \lceil \Phi \right \rceil+\frac{1}{2} \right )$ to $\left \lceil \Phi \right \rceil+\frac{1}{2}$ incrementing in steps of 0.5, which represents loops over the real and imaginary part of the discontinuities. Even though an excess of points is generated for $\Psi$, this method is an efficient way of implementing this part of the algorithm for small number of users. The reason is that since every element of $\Psi$ is used $L$ times to generate elements of $\Omega$, then for small $L$ the excess of points generated for $\Omega$ is also small. For larger values of $L$,  using this imprecise method will lead to a larger amount of excess points for $\Omega$. Therefore it might be more computationally efficient to generate the exact values of $\Psi$ without any additional excess points. One way to generate the exact points of $\Psi$ is by iterating over the same two loops but adjusting the range of values searched in one loop based on the value selected in the other loop. For example, denote $\Re$ as the value selected by the loop generating real part of the discontinuity, and denote $\Im$ as the value generated by the loop representing imaginary values. We iterate the real value loop from  $-\left ( \left \lceil \Phi \right \rceil+\frac{1}{2} \right )$ to $\left \lceil \Phi \right \rceil+\frac{1}{2}$ incrementing in steps of 0.5, and for each value of the real loop $\Re$, we iterate the imaginary value loop from $-\left \lceil \sqrt{\left ( \left \lceil \Phi \right \rceil+\frac{1}{2} \right )^{2}-\Re^{2}} \right \rceil$ to $\left \lceil \sqrt{\left ( \left \lceil \Phi \right \rceil+\frac{1}{2} \right )^{2}-\Re^{2}} \right \rceil$ in steps of 0.5. Even though this leads to additional computational costs to generate the relevant values of $\Psi$, the overall complexity of generating $\Omega$ and subsequent operations of evaluating $\mathbf{a}\left ( \alpha  \right )$ will be lower for larger values of $L$.

Our proposed algorithm can also be made more efficient by recognising that only one quadrant of $\Omega$ needs to be evaluated to completely solve the minimization of $f\left ( \mathbf{a}\left ( \alpha  \right ) \right )$. This is because discontinuities represented by $\Omega$ has a rotational symmetry of four, and Gaussian integer space also has a rotational symmetry order of four. This means any Gaussian integer will become another Gaussian integer of the same modulus after rotation through each of the four quadrants. Therefore the values of $\left \|   \mathbf{a} \right \|$, $\left | \mathbf{a}\mathbf{h}^{T} \right |$ in (\ref{Eq8}), and the overall computational rate of C-F, do not change for any single element of $\Omega$ as it is rotated through each of the four quadrants.  As a result, when evaluating $f\left (\mathbf{a}\left ( \alpha \right ) \right )$ for elements of the $\Omega$, only discontinuities within a single quadrant of $\Omega$ needs to be checked. The proposed method is explicitly outlined in  Algorithm~\ref{alg:alg_1}.

\begin{algorithm*}
\caption{Optimal Coefficient Search Algorithm for C-F over Complex Channels}\label{euclid}
\label{alg:alg_1}
\begin{algorithmic}[1]
\Require  Channel vector $\mathbf{h}$, transmission power $P$
\Ensure Optimal coefficient vector  $\mathbf{a}^{\textup{opt}}$ 
\Algphase{Initialization}
\State $\mathbf{M}=\left (1+P\left \| \mathbf{h} \right \|^{2}  \right )\cdot \mathbf{I}-P\cdot \mathbf{h}^{H}\mathbf{h}$
\State $f\left ( \mathbf{a} \right )=\mathbf{a}\mathbf{M}\mathbf{a}^{H}$
\State $f_{\textup{min}}=\infty$
\State $\Phi =\sqrt{1+P\left \| \mathbf{h} \right \|^{2}}$
\State For $\mathbf{a}\in\mathbb{Z}\left [ j \right ]^{L}$: Define complex set $\Psi :\left \{c+dj, e+fj, o+pj \right \}$, $c\in \mathbb{Z}$, $d-\frac{1}{2}\in \mathbb{Z}$, $e-\frac{1}{2}\in \mathbb{Z}$, $f\in \mathbb{Z}$,  $o-\frac{1}{2}\in \mathbb{Z}$, $p-\frac{1}{2}\in \mathbb{Z}$ where any
 $\varphi \in \Psi $ satisfies $\left \| \mathbf{\varphi } \right \|\leq \left \lceil \Phi \right \rceil+\frac{1}{2}$.
\Statex For $\mathbf{a}\in \mathbb{Z}\left [ \omega \right ]^{L}$: Define complex set $\Psi :\left \{ c+dj, e+fj, o+pj \right \}$ where $c\in \frac{1}{4}+\frac{1}{2}\mathbb{Z}$, $d\in \frac{\sqrt{3}}{4}+\frac{\sqrt{3}}{2}\mathbb{Z}$, $e\in \mathbb{Z}$, $f\in \frac{\sqrt{3}}{2}+\sqrt{3}\mathbb{Z}$, $o\in \frac{1}{2}+\mathbb{Z}$, $p\in\sqrt{3}\mathbb{Z}$  where any $\varphi \in \Psi $ satisifes $\left \| \mathbf{\varphi } \right \|\leq \left \lceil \Phi \right \rceil+\frac{3}{4}$.
\State Define $\mathbb{Z}\left [ j \right ]$ quantizer $\left [ \cdot  \right ]_{\mathbb{Z}\left [ j  \right ]}$ as in Section I: A
\State Define $\mathbb{Z}\left [ \omega  \right ]$ quantizer $\left [ \cdot  \right ]_{\mathbb{Z}\left [ \omega  \right ]}$ as in Section III: B

\Algphase{Phase 1 - Obtain discontinuities of $f\left ( \mathbf{a}\left ( \alpha  \right ) \right )$}
\For{all $l\in \left \{ 1,...,L \right \}$ and $h_{l}\neq 0$}
\For{all $\varphi \in \Psi$}  
\State Calculate $\mathbf{a}=\left [\frac{\varphi }{h_{l}}\mathbf{h} \right ]_{\mathbb{Z}\left [ j \right ]}$ for $\mathbf{a}\in \mathbb{Z}\left [ j \right ]^{L}$ or
\Statex \qquad \quad \qquad  \qquad $\mathbf{a}=\left [\frac{\varphi }{h_{l}}\mathbf{h} \right ]_{\mathbb{Z}\left [ \omega  \right ]}$ for  $\mathbf{a}\in \mathbb{Z}\left [ \omega \right ]^{L}$
\State Calculate  $f\left ( \mathbf{a} \right )=\mathbf{a}\mathbf{M}\mathbf{a}^{H}$
\If{$f\left ( \mathbf{a} \right )< f_{\textup{min}}$ and $\mathbf{a}$ is not an all zero vector}
\State Set $\mathbf{a}^{\textup{opt}}=\mathbf{a}$
\State Set $ f_{\textup{min}}=f\left ( \mathbf{a} \right )$
\EndIf
      \EndFor
\EndFor

\Algphase{Phase 2 -Check unit norm vectors}
\For {all $\mathbf{a} \in  \left \{  \mathbb{Z}\left [ j \right ]^{L} \textup{for search in} \, \mathbb{Z}\left [ j \right ], \,\textup{or} \,\, \mathbb{Z}\left [ \omega \right ]^{L} \textup{for search in} \, \mathbb{Z}\left [ \omega \right ]\right \}$ \,$\textup{where} \, \left \| \mathbf{a} \right \|=1 $}
\If{$f\left ( \mathbf{a}\right )< f_{\textup{min}}$}
\State Set $\mathbf{a}^{\textup{opt}}=\mathbf{a}$
\EndIf
\EndFor
\State  \textbf{return} $\mathbf{a}^{\textup{opt}}$\

\end{algorithmic}
\end{algorithm*}

\subsection{Proposed Algorithm: C-F over Eisenstein Integers}

Eisenstein integers are complex numbers of the form $a+b\omega$ where $a,b\in \mathbb{Z}$ and $\omega =-\frac{1}{2}+j\frac{\sqrt{3}}{2}$. In the case where the relay would like to decode an Eisenstein integer linear combination of codewords with coefficients $\mathbf{a}\in \mathbb{Z}\left [ \omega  \right ]^{L}$, our proposed method can also be adapted to find the optimal Eisenstein integer coefficients, effectively finding the exact solution for,
\begin{equation}\label{Eq23}
\mathbf{a}^{\textup{opt}}=\underset{\mathbf{a}\in \mathbb{Z}[\omega]^{L}\setminus \mathbf{a}= \mathbf{0}}{\operatorname{argmin}}f\left ( \mathbf{a} \right )=\underset{\mathbf{a}\in \mathbb{Z}[\omega]^{L}\setminus \mathbf{a}= \mathbf{0}}{\operatorname{argmin}}\left(\textbf{a}\mathbf{M}\mathbf{a}^{H}  \right ).
\end{equation}

\noindent The proposed optimal coefficient search algorithm for C-F over Eisenstein integers therefore follows the same structure as for Gaussian integers. First we calculate the discontinuities $\mathbf{\varphi }$ between neighbouring Eisenstein integers that satisfy the constraint,
\begin{equation}\label{Eq24}
\left | \mathbf{\varphi } \right |\leq \left \lceil \Phi \right \rceil+\frac{3}{4}.
\end{equation}
\noindent The discontinuities between Eisenstein integers are intermediate points between one Eisenstein integer and a neighbouring Eisenstein integer that is closer to it in Euclidean distance than any other Eisenstein integer. The significance of the discontinuities is that they represent the exact cut-off point for lattice decoding to one Eisenstein integer or to a neighbouring Eisenstein integer. Therefore the discontinuities represent the transition points where the value of $f\left ( \mathbf{a}\left ( \alpha  \right ) \right )$ changes. The discontinuities $\varphi$ between neighbouring Eisenstein integer points that we must consider are,
\begin{equation}\label{Eq25}
\varphi \in \Psi : \begin{cases}
 & c+dj \; \textup{where} \;c\in \frac{1}{4}+\frac{1}{2}\mathbb{Z}, \:d\in \frac{\sqrt{3}}{4}+\frac{\sqrt{3}}{2}\mathbb{Z},\\
 & e+fj \; \textup{where} \;e\in \mathbb{Z}, \:f\in \frac{\sqrt{3}}{2}+\sqrt{3}\mathbb{Z},\\
 & o+pj \;\textup{where} \; o\in \frac{1}{2}+\mathbb{Z}, \:p\in\sqrt{3}\mathbb{Z}.
\end{cases}
\end{equation}

\noindent Note that each of the three types of discontinuites in (\ref{Eq25}) can be generated by iterating over two loops for the real and imaginary components of the discontinuities. Based on (\ref{Eq24}), the range of each loop will be $-\left (\left \lceil \Phi \right \rceil+\frac{3}{4}  \right )$ to $\left \lceil \Phi \right \rceil+\frac{3}{4} $.

Using the discontinuities of $\mathbf{a}\left ( \alpha  \right )\in \mathbb{Z}\left [ \omega  \right ]^{L}$ denoted by $\varphi$ and the relationship specified in Theorem 1, we can obtain the values of $\alpha$ for which $f\left (\mathbf{a}\left ( \alpha  \right )  \right )$ is discontinuous. $\Omega$ in (\ref{Eq26}) denotes the set of all discontinuities of $\alpha$ that must be evaluated,
\begin{equation}\label{Eq26}
\Omega =\left \{\frac{\varphi }{h_{l}}\mid l=1,...,L, \, h_{l}\neq 0, \,  \left | \mathbf{\varphi } \right |\leq \left \lceil \Phi \right \rceil+\frac{3}{4}  \right \}.
\end{equation}

\noindent Using each discontinuity point in $\Omega$ we calculate a corresponding Eisenstein integer coefficient vector using Theorem 1, taking the optimal Eisenstein integer coefficient to be the  $\mathbf{a}\left ( \alpha  \right )\in \mathbb{Z}\left [ \omega  \right ]^{L}$ that minimises the target cost function (\ref{Eq23}). A number of efficient $A_{2}$ hexagonal lattice decoders are proposed in \cite{conway1982fast} and \cite{mow1992fast}, which can be used to perform $\left [ \cdot  \right ]_{\mathbb{Z}\left [ \omega  \right ]}$, the quantization operation over Eisenstein integers. Since the efficiency of the $A_{2}$ decoder will have a significant impact on the overall search speed of our algorithm, the most efficient $A_{2}$ decoder should be used. In\cite{mow1992fast}, a very efficient $A_{2}$ decoder is proposed by breaking down the $A_{2}$ lattice as a union of a rectangular lattice $B=\left \{\left ( u, v \right ) ;u\in \mathbb{Z},v\in \sqrt{3}\mathbb{Z} \right \}$ and its coset $B+ \left ( \frac{1}{2},\frac{\sqrt{3}}{2} \right )$ giving the lattice $A_{2}=B\cup \left (B+ \left ( \frac{1}{2},\frac{\sqrt{3}}{2} \right )\right )$.  Let $x\in \mathbb{R}^{2}$ be the input to the decoder. Let $Q_{B}\left ( x \right )$ be the closest point of $x$ in $B$. To decode $x$ to the nearest point in the $A_{2}$ lattice, first perform $Q_{B}\left ( x \right )$ to decode to the closest point in the rectangular lattice $B$, and $Q_{B} \left ( x-\left ( \frac{1}{2},\frac{\sqrt{3}}{2} \right ) \right )+\left ( \frac{1}{2},\frac{\sqrt{3}}{2} \right )$ to decode to the nearest point on its coset $B+ \left ( \frac{1}{2},\frac{\sqrt{3}}{2} \right )$. Given the output from the two rectangular lattice decoders, the lattice point with the minimum Euclidean distance to $x$ is the output for the $A_{2}$ decoder. In the case of a tie, the lattice point with the largest norm is selected. The proposed search method over Eisenstein integers is included as part of Algorithm 1.

\subsection{Algorithm for MIMO C-F over Complex channels}

\begin{algorithm*}
\caption{Optimal Coefficient Search Algorithm for MIMO C-F over Complex Channels}\label{euclid}
\label{alg:alg_2}
\begin{algorithmic}[1]
\Require Channel matrix $\mathbf{H}$, transmission power $P$, Gram matrix $\mathbf{M}=\mathbf{V}\mathbf{D}\mathbf{V}^{H}$ as defined.
\Ensure Optimal coefficient vectors $\mathbf{a}^{\textup{opt}}$.

\Algphase{Initialization}
\State $\lambda_{\textup{max}}$ is the maximum singular value of $\mathbf{H}$
\State $f_{\textup{min}}=\infty$
\State $\Phi =\sqrt{1+P\lambda_{max}^{2}}$
\State For $\mathbf{a}\in \mathbb{Z}\left [ j \right ]^{L}$: Define complex set $\Psi :\left \{c+dj, e+fj, o+pj \right \}$, $c\in \mathbb{Z}$, $d-\frac{1}{2}\in \mathbb{Z}$, $e-\frac{1}{2}\in \mathbb{Z}$, $f\in \mathbb{Z}$,  $o-\frac{1}{2}\in \mathbb{Z}$, $p-\frac{1}{2}\in \mathbb{Z}$ where any
 $\varphi \in \Psi $ satsifies $\left \| \mathbf{\varphi } \right \|\leq \left \lceil \Phi \right \rceil+\frac{1}{2}$.
\Statex For $\mathbf{a}\in \mathbb{Z}\left [ \omega \right ]^{L}$: Define complex set $\Psi :\left \{ c+dj, e+fj, o+pj \right \}$ where $c\in \frac{1}{4}+\frac{1}{2}\mathbb{Z}$, $d\in \frac{\sqrt{3}}{4}+\frac{\sqrt{3}}{2}\mathbb{Z}$, $e\in \mathbb{Z}$, $f\in \frac{\sqrt{3}}{2}+\sqrt{3}\mathbb{Z}$, $o\in \frac{1}{2}+\mathbb{Z}$, $p\in\sqrt{3}\mathbb{Z}$  where any $\varphi \in \Psi $ satisifes $\left \| \mathbf{\varphi } \right \|\leq \left \lceil \Phi \right \rceil+\frac{3}{4}$.
\State Define $\mathbb{Z}\left [ j \right ]$ quantizer $\left [ \cdot  \right ]_{\mathbb{Z}\left [ j  \right ]}$ as in Section I: A
\State Define ${\mathbb{Z}\left [ \omega  \right ]}$ quantizer $\left [ \cdot  \right ]_{\mathbb{Z}\left [ \omega  \right ]}$ as in Section III: B

\Algphase{Phase 1 - Obtain discontinuities of $f\left ( \mathbf{a}\left ( \alpha  \right ) \right )$}
\For{all $ \tau \subseteq \left \{ 1,...,L \right \}, \left |\tau \right |=k, \left ( \mathbf{H}\right )_{\tau} \,\textup{full rank}$}
\For{all $\mathbf{c}\in \Psi^{k}$, $ \left |\mathbf{c}  \right |\leq \begin{cases}
\left (    \left \lceil \Phi \right \rceil+\frac{1}{2} \right )\mathbf{1} & \text{ for }  \mathbf{a}\in \mathbb{Z}\left [ j \right ]^{L}\\ 
\left (    \left \lceil \Phi \right \rceil+\frac{3}{4} \right )\mathbf{1} & \text{ for } \  \mathbf{a}\in \mathbb{Z}\left [ \omega \right ]^{L}
\end{cases}$} 
\State Calculate $\mathbf{a}=\left [\mathbf{c} \left ( \mathbf{H}_{\tau}  \right )^{-1}\mathbf{H}\right ]_{\mathbb{Z}\left [ j\right ]} $ for $\mathbf{a}\in \mathbb{Z}\left [ j \right ]$, or $\mathbf{a}=\left [\mathbf{c}\left ( \mathbf{H}_{\tau}  \right )^{-1}\mathbf{H}\right ]_{\mathbb{Z}\left [ \omega \right ]}$ for $\mathbf{a}\in \mathbb{Z}\left [ \omega \right ]$
\If{$f\left ( \mathbf{a} \right )=\mathbf{a}\mathbf{M}\mathbf{a}^{H}< f_{\textup{min}}$ and $\mathbf{a}$ is not a zero vector}
\State Set $\mathbf{a}^{\textup{opt}}=\mathbf{a}$
\State Set $ f_{\textup{min}}=f\left ( \mathbf{a} \right )$
\EndIf
      \EndFor
\EndFor

\Algphase{Phase 2 -Check unit norm vectors}
\For {all $\mathbf{a} \in  \left \{  \mathbb{Z}\left [ j \right ]^{L} \textup{for search in} \, \mathbb{Z}\left [ j \right ], \,\textup{or} \,\, \mathbb{Z}\left [ \omega \right ]^{L} \textup{for search in} \, \mathbb{Z}\left [ \omega \right ]\right \}$ \,$\textup{where} \, \left \| \mathbf{a} \right \|=1 $} 
\If{$f\left ( \mathbf{a} \right )< f_{\textup{min}}$}
\State Set $\mathbf{a}^{\textup{opt}}=\mathbf{a}$
\EndIf
\EndFor
\State  \textbf{return} $\mathbf{a}^{\textup{opt}}$

\end{algorithmic}
\end{algorithm*}

Using Theorem 1, we adapt our proposed search algorithm to MIMO C-F setting over complex channels, where the minimization function in (\ref{Eq15}) can be again represented as a multidimensional function with constant polytope regions that must be checked. However, for $k$ number of receive antennas, each vertex is now the intersection of $k$ linearly independent hyperplanes of the form $c_{n}=\mathbf{b}\mathbf{h}_{n}^{T}$ where  $c_{n}\in \Psi$  is the Gaussian integer discontinuity defined in (\ref{Eq19}). Note that for optimal coefficient vector search over Eisenstein integers, simply set $c_{n}\in \Psi$ as in (\ref{Eq25}). Therefore, to obtain all the vertex points of the cost function $f\left ( \mathbf{a}\left ( \alpha  \right ) \right )$, choose any set $\tau \subseteq \left \{ 1,...,L \right \}$ for which $\left |\tau  \right |=k $, with $\mathbf{H}_{\tau }$ full ranked, and solve $\mathbf{c}=\mathbf{b} \mathbf{H} _{\tau }$ for $\mathbf{b}$ where vector $\mathbf{c}$'s individual elements take values from the set of discontinuities $\varphi$ in (\ref{Eq19}) and (\ref{Eq25}) for search over Gaussian integers and Eisenstein integers respectively. 

Based on the range for optimal coefficient vector for MIMO C-F given by (\ref{Eq14}), it's clear that only vertices within a specific range have to be considered,
\begin{equation}
\left |\mathbf{c}  \right |\leq \begin{cases}\label{Eq27}
 \left (    \left \lceil \Phi \right \rceil+\frac{1}{2} \right )\mathbf{1} & \text{ for } \textup{search over}\  \mathbf{a}\in \mathbb{Z}\left [ j \right ],\\ 
\left (    \left \lceil \Phi \right \rceil+\frac{3}{4} \right )\mathbf{1} & \text{ for } \textup{search over}\  \mathbf{a}\in \mathbb{Z}\left [ \omega \right ].
\end{cases}
\end{equation}
Therefore for C-F over Gaussian integers, the discontinuities of $f\left ( \mathbf{a}\left ( \alpha  \right ) \right )$ that must be checked are all points in the set $\Omega _{\tau }$,
\begin{equation}\label{Eq28}
\begin{split}
\Omega _{\tau }&=   \left \{     \mathbf{c}  \left ( \mathbf{H}_{\tau}\right )^{-1} \mid   \left |\mathbf{c}  \right |\leq \left (    \left \lceil \Phi \right \rceil+\frac{1}{2} \right )\mathbf{1}, \, \mathbf{c}\in \Psi^{k}
  \right \}.\\
& \tau \subseteq \left \{ 1,...,L \right \}, \left |\tau  \right |=k, \, \mathbf{H}_{\tau} \,\textup{full rank}\\
& \Omega = \bigcup_{ \tau\subseteq \left \{ 1, ..., L \right \}, \, \left | \tau \right |=k}\Omega_{\tau} 
\end{split}
\end{equation}

\noindent For optimal search over Eisenstein integers, the relevant bound in (\ref{Eq24}) should be used to evaluate the relevant discontinuities of interest. Finally using Theorem 1, each vertex obtained in $\Omega _{\tau }$ can be used directly to calculate an C-F coefficient vector either over Gaussian integers or Eisenstein integers. These coefficient vectors can then be used to  evaluate cost function $f\left ( \mathbf{a}\left ( \alpha  \right ) \right )$ without missing any potential minimal regions. The search method over MIMO C-F is outlined explicitly in Algorithm 2.

\subsection{Complexity Analysis}

In this section, the complexity of Algorithm~\ref{alg:alg_1} and Algorithm~\ref{alg:alg_2} are analysed. For Algorithm~\ref{alg:alg_1}, the complexity of the algorithm over both Gaussian integers and Eisenstein integers are provided. For Algorithm 2 over MIMO C-F, the complexity of search over Gaussian integers is given while its complexity over Eisenstein integers can be deduced in a similar manner. 

To find the optimal coefficients over Gaussian integers, the first part of Algorithm~\ref{alg:alg_1} consists of finding all the $\alpha$ values for which  $f\left ( \mathbf{a}\left ( \alpha  \right ) \right )$ is discontinuous, the complexity of which depends on the number of discontinuities of $\mathbf{a}\left ( \alpha  \right )$ that satisfies the bound in (\ref{Eq20}). This is equivalent to finding the number of discontinuities of $\mathbf{a}\left ( \alpha  \right )$ that lie within a sphere of radius $\left \lceil \Phi \right \rceil+\frac{1}{2}$, which is a form of Gauss circle problem. We can suggest a loose upperbound for the number of discontinuities to be calculated using the ratio of area of a sphere of radius $\left \lceil \Phi \right \rceil+\frac{1}{2}$, relative to a square with sides of length $2\left \lceil \Phi \right \rceil+1$. Within the bound defined in (\ref{Eq20}), in each axis, there are $2\lceil \Phi \rceil$ number of integers and $2\lceil \Phi \rceil+2$  number of half integers. Therefore within a square of sides  $2\left \lceil \Phi \right \rceil+1$ centred at the origin, there are a total of $12\left \lceil \Phi  \right \rceil^{2}+12\left \lceil \Phi  \right \rceil+4$ discontinuities of $\mathbf{a}\left ( \alpha \right )$ of the form given in (\ref{Eq19}). The area ratio between a sphere of radius $\left \lceil \Phi \right \rceil+\frac{1}{2}$ and a square of with sides of length  $2\left \lceil \Phi \right \rceil+1$, is given by $\pi /4$. Therefore we can deduce the upper bound on the total number of discontinuities of $\mathbf{a}\left ( \alpha \right )$  to be searched as,
\begin{equation*}\label{Eq28}
\lceil  \pi   \left(  3\lceil \Phi  \rceil^{2}+3\lceil \Phi \rceil+1 \right )\rceil.
\end{equation*}

\noindent Taking into account the number of transmitting users $L$, the total number points of $\alpha$ for which $f\left ( \mathbf{a}\left ( \alpha  \right ) \right )$ is discontinuous is given by,
\begin{equation*}\label{Eq30}
L\lceil  \pi   \left(  3\lceil \Phi  \rceil^{2}+3\lceil \Phi \rceil+1 \right )\rceil.
\end{equation*}

\noindent Since we only need to check discontinuities in one of the quadrants of the complex set $\Omega$ to evaluate $f\left ( \mathbf{a}\left ( \alpha  \right ) \right )$ completely, the complexity is quartered,
\begin{equation*}
0.25L\lceil  \pi   \left(  3\lceil \Phi  \rceil^{2}+3\lceil \Phi \rceil+1 \right )\rceil.
\end{equation*}
\noindent Finally to enumerate the value of $f\left ( \mathbf{a} \right )=\mathbf{a}\mathbf{M}\mathbf{a}^{H}$, $L$ number of calculations are performed to obtain the summation resulting from the matrix multiplication. The overall complexity of Algorithm~\ref{alg:alg_1} for the complex channel case is therefore given by, 
\begin{equation*}
0.25L^{2}\lceil  \pi   \left(  3\lceil \Phi  \rceil^{2}+3\lceil \Phi \rceil+1 \right )\rceil.
\end{equation*}

The complexity of Algorithm~\ref{alg:alg_1} for finding the optimal C-F coefficients over Eisenstein integers can be characterised by three main parts. The complexity of the first part derives from obtaining all the discontinuities of $f\left (\mathbf{a}\left ( \alpha  \right )  \right )$ defined in (\ref{Eq25}). Following the same method of analysis for Gaussian integers, for $L$ users it can be shown that the number of discontinuities of $f\left (\mathbf{a}\left ( \alpha  \right )  \right )$ is of $O\left (L \Phi^{2} \right )$. In the second part of the algorithm, each $\alpha$ discontinuity is used to obtain a corresponding $\mathbf{a}\left ( \alpha  \right )$ via the $A_{2}$ decoder which means the speed of the overall algorithm is also highly dependent on the efficiency of the $A_{2}$ decoder. Finally the value of $f\left (\mathbf{a}\left ( \alpha  \right )  \right )$  can be calculated in $O\left (L  \right )$. The overall complexity of the algorithm is therefore given by  $O\left ( L^{2} \Phi ^{2} \right )$. In comparison to Algorithm~\ref{alg:alg_1} over Gaussian integers, the search over Eisenstein integers runs slightly slower in part due to the need to check more discontinuities, and also due to use of the $A_{2}$ decoder to  generate the quantized coefficients.

For complexity of Algorithm 2 for MIMO C-F, there are approximately $(\lceil  \pi   \left(  3\lceil \Phi  \rceil^{2}+3\lceil \Phi \rceil+1 \right )\rceil)^{k}$ discontinuities of $\mathbf{a}\left ( \alpha \right )\in \mathbb{Z}\left [ j \right ]^{L}$ satisfying bound  (\ref{Eq20}). For each of these discontinuities, $\mathbf{c}\left ( \mathbf{H}_{\tau}  \right )^{-1}$ is calculated $P\left ( L,k \right )$ times to obtain all the discontinuities of $f\left ( \mathbf{a} \left ( \alpha \right )\right )$ in $\alpha$,
\begin{equation*}
\binom{L}{k}(\lceil  \pi   \left(  3\lceil \Phi  \rceil^{2}+3\lceil \Phi \rceil+1 \right )\rceil)^{k}.
\end{equation*}
This expression can be upperbounded by, 
\begin{equation*} 
\begin{split}
&\frac{L^{k}}{k!}(\lceil  \pi   \left(  3\lceil \Phi  \rceil^{2}+3\lceil \Phi \rceil+1 \right )\rceil)^{k}\\
&=\frac{(L\lceil  \pi   \left(  3\lceil \Phi  \rceil^{2}+3\lceil \Phi \rceil+1 \right )\rceil)^{k}}{k!}.
\end{split}
\end{equation*}

\noindent Each discontinuity of $f\left ( \mathbf{a} \left ( \alpha \right )\right )$ is used to calculate $\mathbf{a}=\left [\mathbf{c}  \left ( \mathbf{H}_{\tau}  \right )^{-1}\mathbf{H}\right ]_{\mathbb{Z}\left [ j \right ]}$ and evaluate $f\left ( \mathbf{a} \left ( \alpha \right )\right )$, both operations can be calculated at $O\left ( kL \right )$. Therefore the complexity of Algorithm 2 can be denoted by, 

\begin{equation*}
\begin{split}
&O\left ( kL\frac{(L\lceil  \pi   \left(  3\lceil \Phi  \rceil^{2}+3\lceil \Phi \rceil+1 \right )\rceil)^{k}}{k!} \right )\\
&=O\left (\frac{L^{k+1}\lceil  \pi   \left(  3\lceil \Phi  \rceil^{2}+3\lceil \Phi \rceil+1 \right )\rceil^{k}}{\left ( k-1 \right )!} \right ).
\end{split}
\end{equation*}

\noindent In \cite{sahraei2014compute}, it is shown that the term $\Phi$ is upperbounded by a polynomial function of $L$. Therefore, the complexity of Algorithm 2 for MIMO C-F is of polynomial complexity in $L$. From an implementation perspective, the search speed of Algorithm 2 will be slower than Algorithm 1 for two main seasons. Firstly with Algorithm 2 the complexity of evaluating total number of permutations for discontinuities $\mathbf{c}\in \Psi^{k}$ that satisfy the bound $\left |\mathbf{c}  \right |\leq \left (    \left \lceil \Phi \right \rceil+\frac{1}{2} \right )\mathbf{1}$ is much higher than for Algorithm 1, where only discontinuties $\varphi \in \Psi $ within the bound $\left | \mathbf{\varphi } \right |\leq \left \lceil \Phi \right \rceil+\frac{3}{4}$ needs to be evaluated. Additionally, the matrix inversion operation $\left ( \mathbf{H}_{\tau}  \right )^{-1}$ which must be performed for each set of discontinuities also increases runtime significantly for each iteration in comparison to Algorithm 1 where only a division by a complex variable is required to obtain the discontinuities of $f\left ( \mathbf{a}\left ( \alpha \right ) \right )$. 

\section{Numerical Results}

We verify the performance of our proposed Algorithm 1 for standard C-F over the ring of Gaussian integers by running simulations to examine its performance and efficiency. All the numerical results are obtained using MATLAB on an Intel Core(TM) i5 CPU @2.2GHz. 

\begin{figure}[htb]
\centering
\includegraphics[width=9cm]{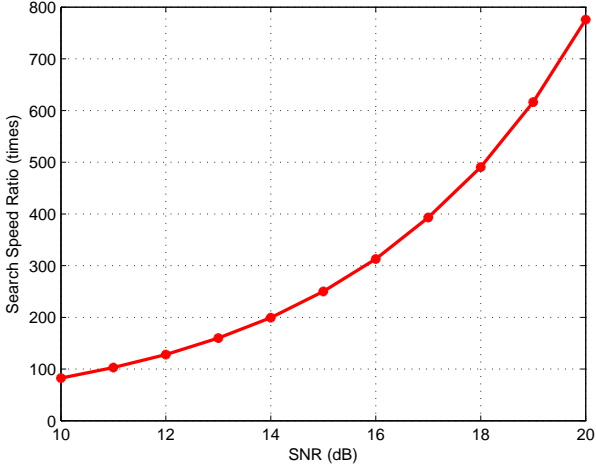}
\caption{Runtime performance ratio $\left (  \frac{\textup{Exhaustive Search CPU time}}{\textup{Algorithm 1 CPU time}}\right )$ against SNR for $L=2$.}
\label{fig:2}
\end{figure}

In Figure~\ref{fig:2} we compare the efficiency of  Algorithm~\ref{alg:alg_1} against the exhaustive search method for C-F over Gaussian integers, with two transmitting users and single receive antenna.  The wireless channel between each user and the relay node is assumed to be complex Gaussian channel. $5000$ channel instances are generated for each SNR. The search duration for each channel instance is accumulated and the total CPU runtime is compared between Algorithm 1 and the exhaustive search method. The ratio of CPU runtime for exhaustive search over the Algorithm 1 is given in Figure~\ref{fig:2}. It can be observed that as SNR increases, our proposed search algorithm runs much faster than the exhaustive search method. This is consistent with theory as for two users, the complexity of our proposed Algorithm 1 increases at the rate of $O(\Phi ^{2})$ as the SNR increases, while the exhaustive search algorithm increases at the rate $O(\Phi ^{4})$. 

\begin{figure}[htb]
\centering
\includegraphics[width=9cm]{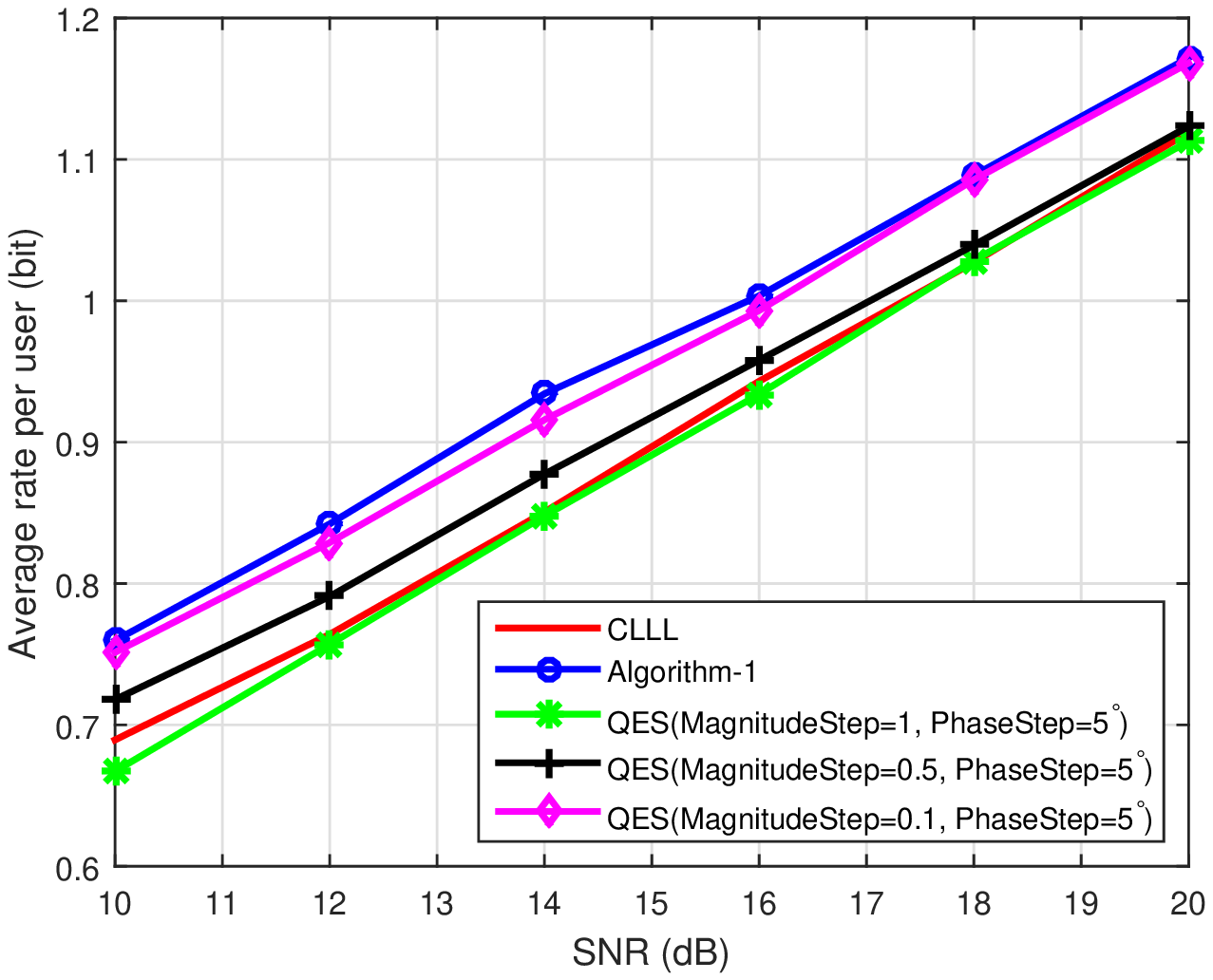}
\caption{Rate performance comparison of Algorithm 1 against CLLL and QES with various search step parameters for $L$=4.}
\label{fig:3}
\end{figure}

\begin{figure}[htb]
\centering
\includegraphics[width=9cm]{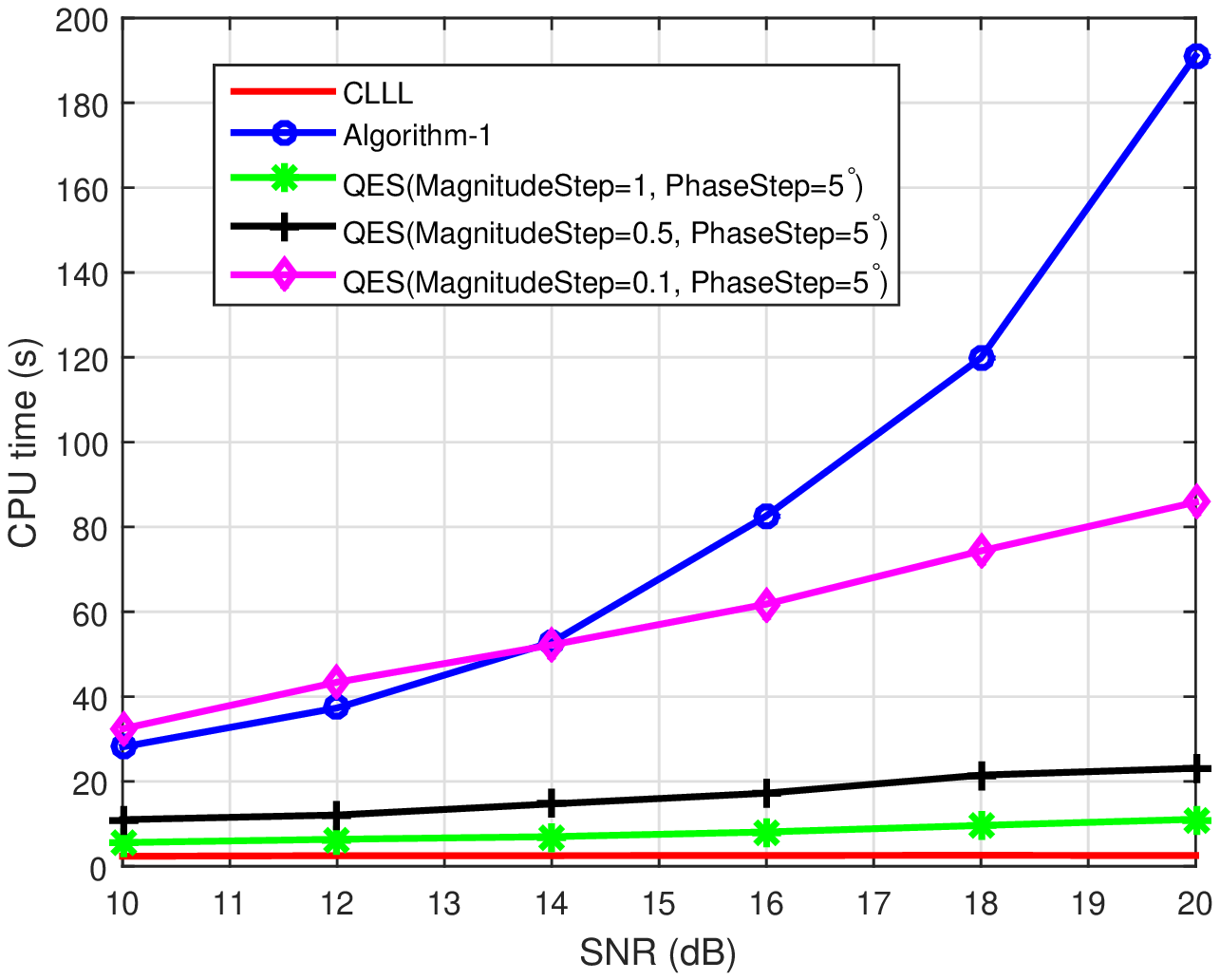}
\caption{CPU runtime comparison of Algorithm 1 against CLLL and QES with various search step parameters for $L$=4.}
\label{fig:4}
\end{figure}

\begin{figure}[htb]
\centering
\includegraphics[width=9cm]{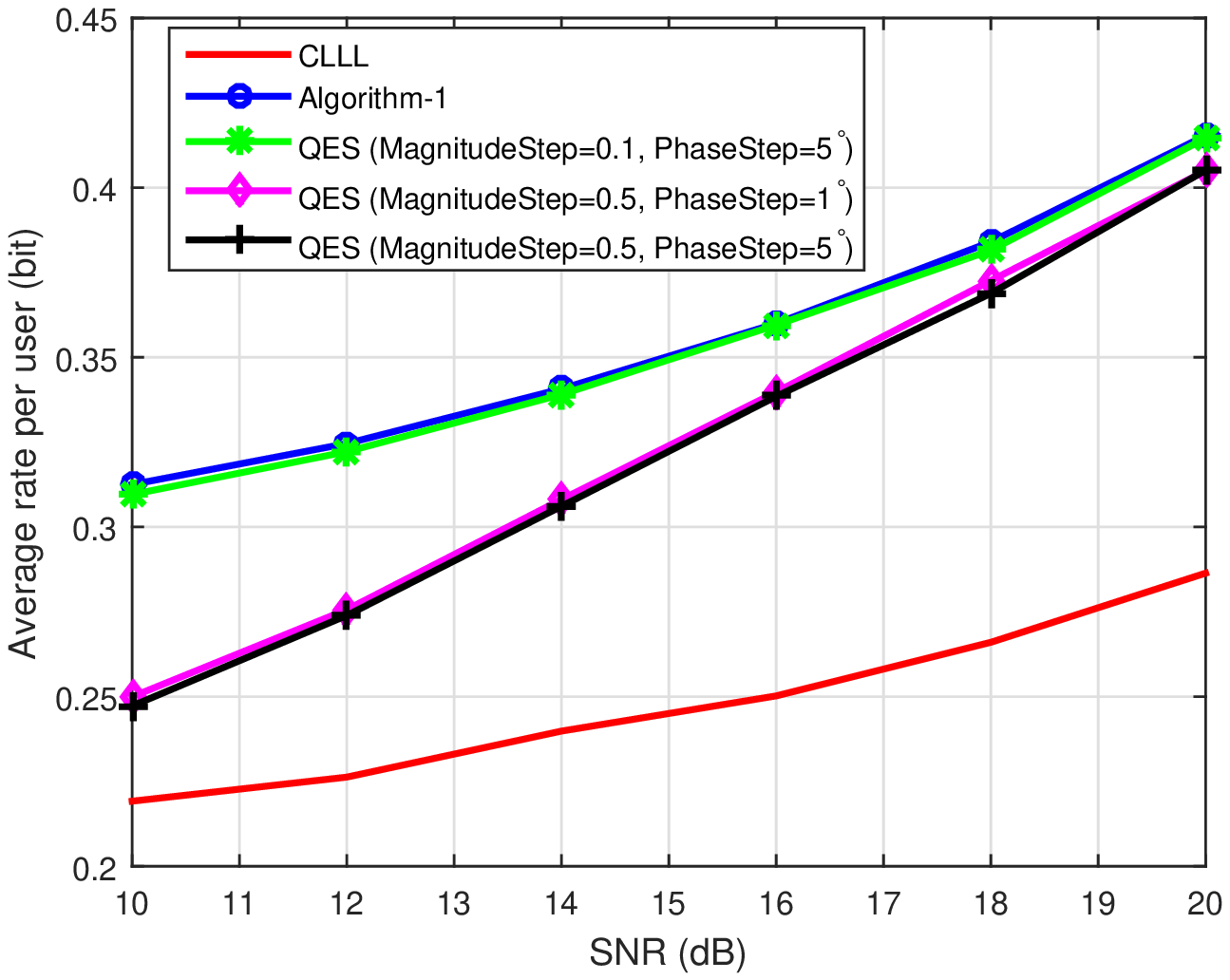}
\caption{Rate performance comparison of Algorithm 1 against CLLL and QES with various search step parameters for $L$=8 }
\label{fig:5}
\end{figure}

\begin{figure}[htb]
\centering
\includegraphics[width=9cm]{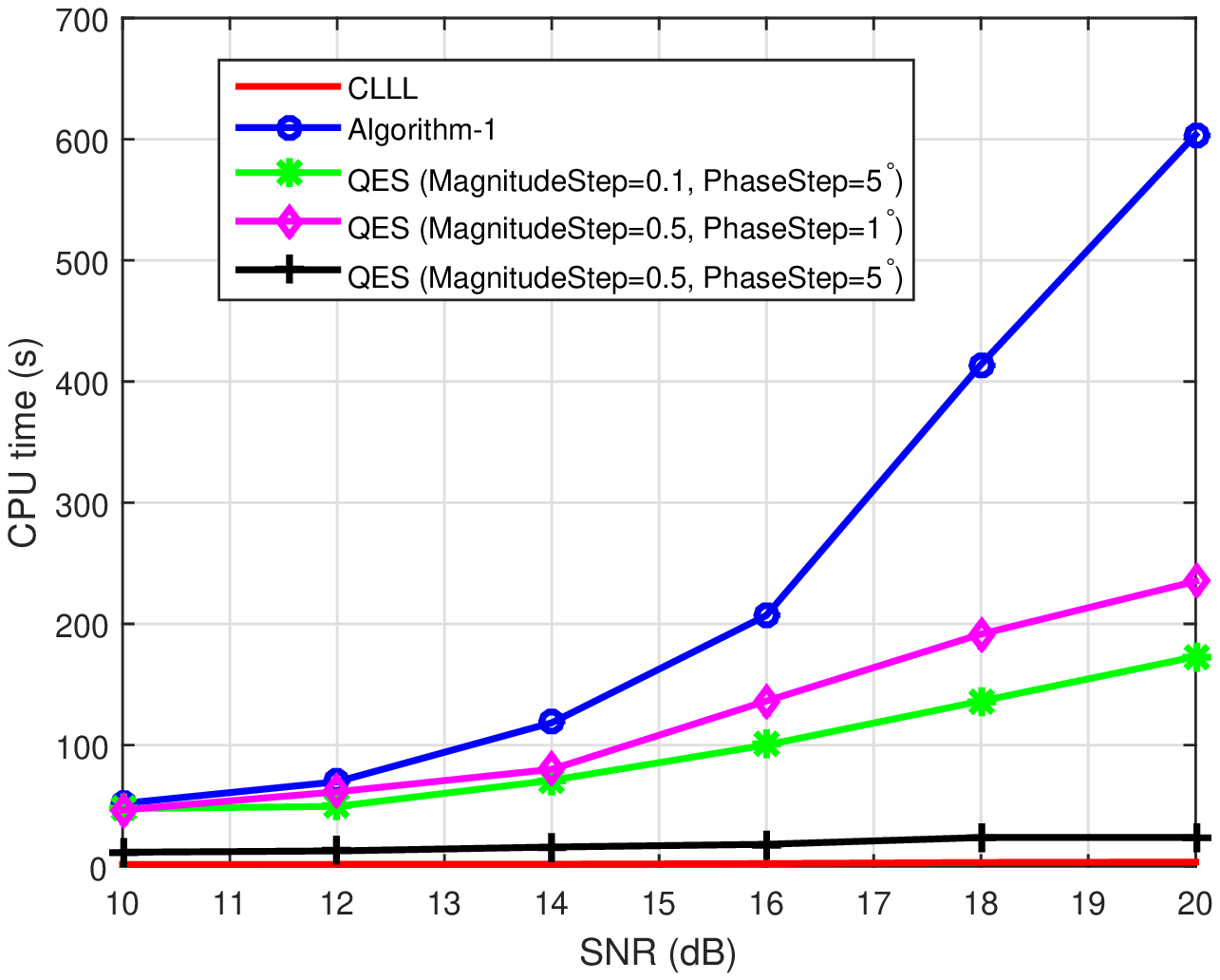}
\caption{CPU runtime comparison of Algorithm 1 against CLLL and QES with various search step parameters for $L$=8.}
\label{fig:6}
\end{figure}

\begin{figure}[htb]
\centering
\includegraphics[width=9cm]{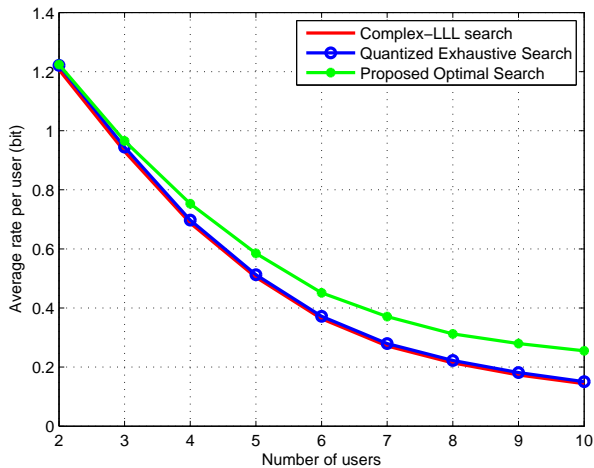}
\caption{Rate performance against transmitting user number $L$, for Algorithm 1, CLLL and QES, at SNR=10dB.}
\label{fig:7}
\end{figure}

\begin{figure}[htb]
\centering
\includegraphics[width=9cm]{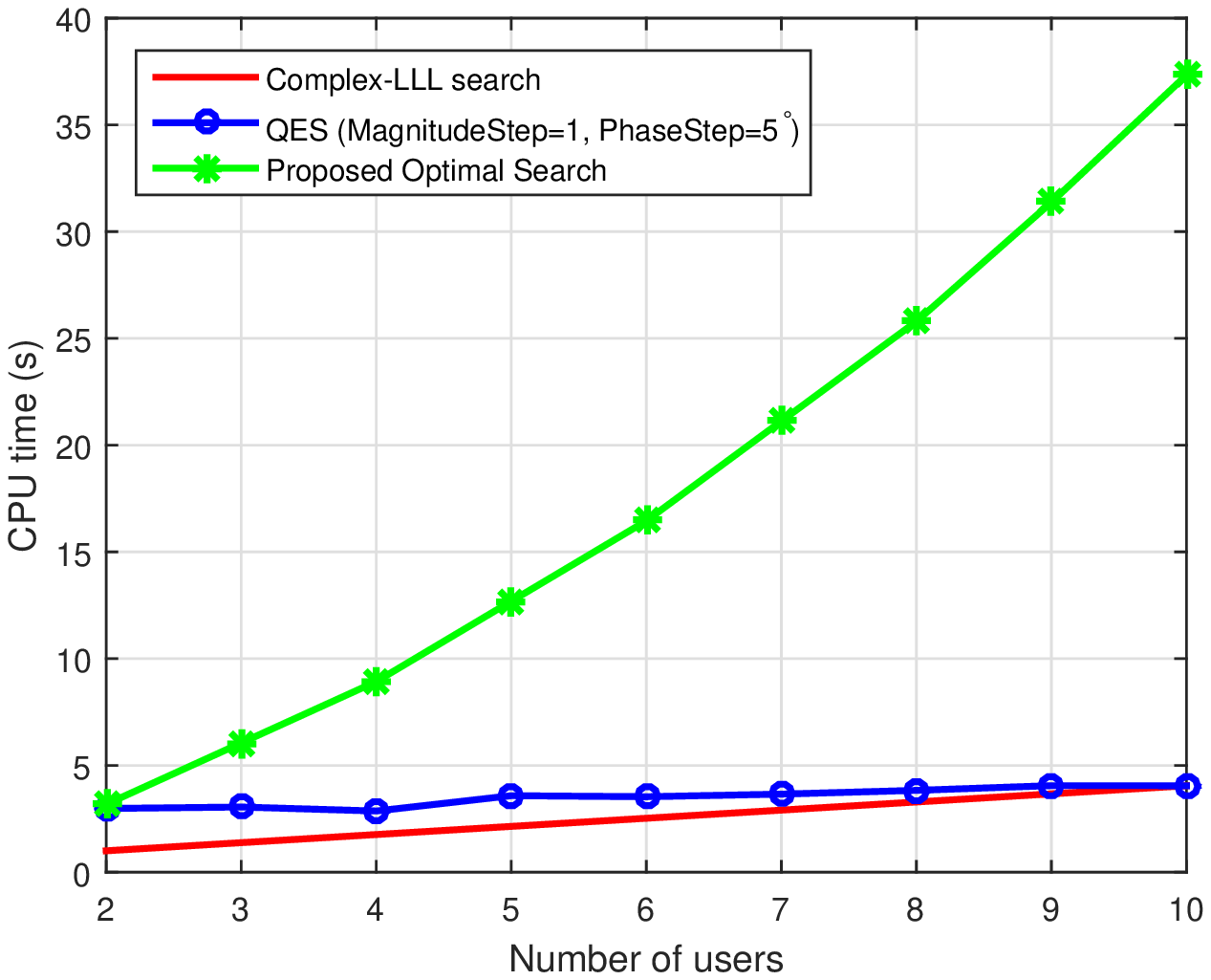}
\caption{CPU runtime against user number $L$ for Algorithm 1, CLLL and QES, at SNR=10dB}
\label{fig:8}
\end{figure}

In Figure~\ref{fig:3} and Figure~\ref{fig:5}, the average C-F rates of  Algorithm~\ref{alg:alg_1}, QES and CLLL are compared for $L=4$ and $L=8$ respectively. The computational rate is taken as rate per user which is obtained by averaging the rate for 5000 channel instances. For QES, simulations with three different sets of  search step parameters are performed. The step sizes are selected to ensure that QES performs at least on par with CLLL in terms of achievable rate, with CLLL taken as the baseline comparison. From Figure~\ref{fig:3} and Figure~\ref{fig:5} it can be seen that  Algorithm~\ref{alg:alg_1} produces the optimal rate while the QES simulations approach optimal rate for finer step size selections. Figure~\ref{fig:4} and Figure~\ref{fig:6} compares the efficiency of the three search algorithms,  Algorithm~\ref{alg:alg_1}, CLLL and 3 versions of QES with different search steps, for $L=4$ and $L=8$ respectively, charting their CPU runtime against SNR. 

In terms of performance against complexity trade-off comparison, we focus on  Algorithm~\ref{alg:alg_1} and QES which are the two most competitive search algorithms investigated in this work. For $L=4$ we observe QES can achieve within 5\% of the optimal rate at a much lower CPU runtime cost than the proposed  Algorithm~\ref{alg:alg_1}, especially for high SNR values. QES(MagnitudeStep= 0.1, PhaseStep=5$^{\circ}$) provides near optimal rate performance to  Algorithm~\ref{alg:alg_1} at slightly higher computational cost than  Algorithm~\ref{alg:alg_1} at lower SNR values. At higher SNR, the efficiency of QES becomes more distinctive. For lower SNR values of the QES(MagnitudeStep=0.1, PhaseStep=5$^{\circ}$) plot, it's worth noting that  even when step sizes are set to a minute level that pushes QES runtime to exceed the runtime of Algorithm 1, the QES will still return a noticeable amount of suboptimal coefficient vectors which is reflected in the slight gap from the optimal rate. Another observation we make based on the data in Figure~\ref{fig:4} is that as we fine tune the QES step size, the runtime increases significantly, yet the improvement in C-F rate diminishes considerably as QES approaches optimality. This can be observed by comparing the rate and CPU time for QES(MagnitudeStep=0.5, PhaseStep=5$^{\circ}$) and QES(MagnitudeStep=0.1, PhaseStep=5$^{\circ}$) where the improvement in rate performance is approximately 5\% across all SNR values, yet the CPU time increased by over 100\% at 10dB and over 300\% at 20dB.

For $L=8$ transmitting users, by comparing the plots for QES(MagnitudeStep=0.5, PhaseStep=5$^{\circ}$) against plot for Algorithm~\ref{alg:alg_1}, for SNR of 10dB it can be seen that  Algorithm~\ref{alg:alg_1} gives approximately 30\% improvement in rate at the cost of 5 times the runtime of QES. At higher SNR values, this trade-off becomes much worse, since the complexity of our proposed method increases exponentially with SNR which is consistent with our analysis of its complexity at $O\left ( L^{2} \Phi ^{2} \right )$. Nevertheless, even using minute QES step values that result in excessively long runtime, QES still returned non-optimal coefficient vectors for a large number of channel instances, as it does for the $L=4$ simulations. QES(MagnitudeStep=0.1,  PhaseStep=5$^{\circ}$) provides near optimal rate performance at slightly lower complexity than Algorithm~\ref{alg:alg_1} for smaller SNR values. At higher SNR, the better efficiency of QES becomes distinctively larger. Comparing QES(MagnitudeStep=0.5, PhaseStep=5$^{\circ}$) with QES(MagnitudeStep=0.5, PhaseStep=1$^{\circ}$) we note that runtime drastically increases with no noticeable change in rate, which again suggests there is a threshold on the extent altering the magnitude and phase steps can efficiently improve the performance of the QES for a specific set of channel parameters. Using the numerical data available, we have found that for certain SNR values it is not possible for QES to return 100\% optimal coefficient vectors, even when its runtime exceeds that of  Algorithm~\ref{alg:alg_1}. Therefore, we conclude that while QES is a highly efficient and competitive search algorithm which can approach optimality at a much smaller computational cost than  Algorithm~\ref{alg:alg_1}, it cannot guarantee to return the exact optimal coefficients for random channel instances, even when search steps are set to miniscule values. Nevertheless, because the number of missed optimal coefficients is so small for detailed QES search steps, the overall loss can be made neglible, thus giving a very good trade-off between performance and complexity.

Figure~\ref{fig:7} illustrates the rate performance of each search method for C-F as $L$ increases while Figure~\ref{fig:8} shows the corresponding CPU runtime.  Note that QES plots in Figure~\ref{fig:7} and Figure~\ref{fig:8} assume a fixed magnitude and phase step size which was selected to ensure QES performed at least on par with CLLL. We observe that the computational rate per user of C-F naturally decreases as $L$ increases due to increased interference between users. The performance of CLLL and QES both deteriorate with increasing $L$ in a similar manner. This is consistent with theory as the performance of CLLL is known to deteriorate at an exponential factor of $2^{\frac{L-1}{2}}$ as $L$ increases. The performance of QES deteriorates with $L$ for a fixed search step since this parameter should be adjusted to be gradually smaller with increasing $L$ in order to maintain optimality of search. How to adaptively select search step parameters to bring about optimal performance for the QES is still an open problem. However, due to the close relationship between QES and Algorithm 1, it can be conjectured that the QES search steps required to ensure optimal performance depends on the angle and Euclidean distance between the two $\alpha$ discontinuities of $f\left ( \mathbf{a}\left ( \alpha  \right ) \right )$ with the minimum Euclidean distance from each other.

Figure~\ref{fig:8} shows the CPU time of the search algorithms as $L$ increases for SNR of 10dB. The runtime of  Algorithm~\ref{alg:alg_1} increases exponentially with $L$ as predicted by its theoretical complexity of $O\left ( L^{2} \Phi ^{2} \right )$. QES shows gradual increase in runtime with increasing $L$ due to the minimization function term (\ref{Eq10}) which must be evaluated at a complexity of $O\left ( L \right )$ for each $\mathbf{a}\left ( \alpha  \right )$ obtained in the iterative search. Finally, even though CLLL has a theoretical complexity of $O\left ( L^{4} \ \textup{Log}L \right )$, based on actual runtime performance, it is still very efficient because of the low computational complexity of each round of calculation.

\section{Conclusion}
In this paper, we proposed algorithms in low polynomial complexity to optimally solve the shortest vector problem that occurs in C-F over complex channels, for both the single receive antenna case and MIMO C-F. The algorithms can be used to obtain the optimal equation coefficients for C-F over the ring of Gaussian integers and the ring of Eisenstein integers. We derived the computational complexity for our proposed algorithms and used simulation results to verify the gap in actual performance between our proposed search method, exhaustive search, quantized exhaustive search, and complex-LLL, which are the main coefficient search algorithms in literature for C-F over complex value channels. It is shown that our proposed search algorithm for standard C-F can achieve optimal performance at competent computational costs. Although quantized exhaustive search can obtain near optimal rates at very efficient computational cost, it cannot guarantee the optimal solution for arbituary channel instances. For future work, the natural extension would be to extend our search algorithms to compute-and-forward over general algebraic rings. This would be particularly useful in conjunction with an adaptive compute-and-forward scheme which chooses the best ring of imaginary quadratic integers to quantize each complex valued channel coefficient, thus achieving better performance than by working strictly with fixed rings. 

\ifCLASSOPTIONcaptionsoff
  \newpage
\fi












\end{document}